\documentclass[10pt,journal,a4paper,final,oneside,twocolumn]{IEEEtran}
\usepackage{cuted}
\usepackage{stfloats}
\usepackage{times} 
\usepackage{graphicx}
\usepackage{cite}
\usepackage{citesort}
\usepackage{color}
\usepackage{psfrag}
\usepackage{subfigure}
\usepackage{amssymb}
\usepackage{epsfig}
\usepackage{pifont}
\usepackage{amsmath}
\usepackage{array}
\usepackage{multicol}
\usepackage[T1]{fontenc}
\usepackage{comment}
\usepackage{amsthm}
\usepackage{indentfirst}
\usepackage{cases}
\usepackage[ruled, lined, linesnumbered, commentsnumbered, longend]{algorithm2e}
\SetKwInOut{KwIn}{Input}
\SetKwInOut{KwOut}{Output}
\SetKwProg{Init}{Initialization}{:}{end}
\SetKwProg{ft}{[The First Stage]}{}{end}
\SetKwProg{st}{[The Second Stage]}{}{end}
\SetKwProg{tt}{[The Third Stage]}{}{end}
\SetKwProg{PM}{[Power allocation among DL MTs]}{}{end}
\SetKwProg{PS}{[Power allocation among subcarriers of individual MTs]}{}{end}
\SetKwProg{SIC}{[SI suppression]}{}{end}
\graphicspath{{figures/}}
\theoremstyle{remark}

\makeatletter
\def\ifundefined{\@ifundefined}
\makeatother 

\begin{document}

\title{Multicarrier-Division Duplex for Solving the Channel Aging Problem in Massive MIMO Systems}

\author{Bohan Li, Lie-Liang Yang, {\em Fellow,
    IEEE}, Robert G. Maunder, {\em Senior Member, IEEE}, Songlin Sun, {\em Senior Member, IEEE}, Pei Xiao, {\em Senior Member, IEEE} \thanks{B. Li, L.-L. Yang and R. Maunder are with the School of Electronics
    and Computer Science, University of Southampton, SO17 1BJ,
    UK. (E-mail:
    bl2n18, lly, rm@ecs.soton.ac.uk,~http://www-mobile.ecs.soton.ac.uk/lly). S. Sun is with the School of Information and Communication Engineering, Beijing University of Posts and Telecommunications (BUPT). P. Xiao is with 5GIC \& 6GIC,
Institute for Communication Systems (ICS), University of Surrey, Guildford
GU2 7XH, UK. (Email: p.xiao@surrey.ac.uk)
  The project
    was supported in part by the EPSRC, UK, under Project EP/P034284/1 and EP/P03456X/1,
    and in part by the Innovate UK project.}}

\maketitle

\begin{abstract}
The separation of training and data transmission as well as the frequent uplink/downlink (UL/DL) switching make time-division duplex (TDD)-based massive multiple-input multiple-output (mMIMO) systems less competent in fast time-varying scenarios due to the resulted severe channel aging. To this end, a multicarrier-division duplex (MDD) mMIMO scheme associated with two types of well-designed frame structures are introduced for combating channel aging when communicating over fast time-varying channels. To compare with TDD, the corresponding frame structures related to 3GPP standards and their variant forms are presented. The MDD-specific general Wiener predictor and decision-directed Wiener predictor are introduced to predict the channel state information, respectively, in the time domain based on UL pilots and in the frequency domain based on the detected UL data, considering the impact of residual self-interference (SI). Moreover, by applying the zero-forcing precoding and maximum ratio combining, the closed-form approximations for the lower bounded rate achieved by TDD and MDD systems over time-varying channels are derived. Our main conclusion from this study is that the MDD, endowed with the capability of full-duplex but less demand on SI cancellation than in-band full-duplex (IBFD), outperforms both the conventional TDD and IBFD in combating channel aging.

\end{abstract}


\IEEEpeerreviewmaketitle

\section{Introduction}\label{section-MDDCEF-intro}

Massive multiple-input multiple-output (mMIMO) has been regarded as one of the most promising technologies for significantly improving the spectral efficiency (SE) in the fifth-generation (5G) wireless systems~\cite{larsson2013massive}. However, due to the large number of antennas equipped at base station (BS), the acquisition of channel state information (CSI) becomes highly challenging. As the result, time-division duplex (TDD) is deemed to be the only feasible mode in practical mMIMO systems to date, which exploits the channel reciprocity to acquire downlink (DL) CSI from uplink (UL) pilots \cite{bjornson2016massive}. As its counterpart, in frequency-division duplex (FDD) systems, the overhead for training/feedback is proportional to the size of the antenna arrays at BS, resulting in less resources for sending desired data in one coherence block \cite{sanguinetti2019toward}. On the other hand, with the growing demand on the mobile services operated in ground vehicles, high-speed trains, unmanned aerial vehicles (UAVs), etc., the performance of TDD systems is no longer dominated by pilot contamination but the channel aging problem. In other words, due to fast time-varying, the channel at the time when it is estimated via UL training may be very different from that at the time when it is applied for DL preprocessing and UL detection. Moreover, the UL/DL switching intervals required by TDD systems exacerbate the situation. This CSI mismatch may lead to significant degradation of the SE of both UL and DL. 

In order to support the robust communications over fast time-varying channels, channel prediction based on the Wiener filter or Kalman filter has been studied. In \cite{truong2013effects}, the authors considered the channel aging problem in mMIMO systems and applied the Wiener predictor (WP) to predict future CSI. The results show that using channel prediction can partially relieve the channel aging effect and render an increase of sum rate. The authors of \cite{papazafeiropoulos2015deterministic} analyzed the effect of WP for time-varying channels on the system performance. However, in these two papers, the one-step WP relies on the latest channels estimated using the UL pilots that are always available in the front of the symbol to be predicted. Unfortunately, in practical TDD systems, UL pilots are unavailable during DL transmission and hence, the accuracy of channel prediction degrades with time, causing the performance degradation. For this sake, in \cite{kashyap2017performance}, the authors considered a more practical scenario, where no UL pilots are sent during the Kalman filter assisted channel prediction for DL transmission. The results show that the Kalman predictor is capable of improving performance when compared with the case without prediction. However, due to the time-varying effect, the achievable DL rate reduces quickly with time as the result of the fact that the predicted CSI becomes less and less accurate. Recently in \cite{yuan2020machine}, a learning-based approach was introduced, showing that it can outperform the conventional non-linear Kalman predictor in dealing with the channel aging problem. Again, the prediction accuracy of CSI deteriorates as the interval between the channel being predicted and the pilot symbols becomes larger.

In contrast to the half-duplex (HD) mode of TDD, in-band full-duplex (IBFD) has the inherent advantage for solving the channel aging problem. This is because in IBFD systems, DL and UL occur concurrently, which enables to acquire the latest CSI during data transmission without invoking DL/UL switching. However, to the best of our knowledge, there are no open references which have considered the IBFD-relied transmission design for the multicarrier mMIMO systems communicating over fast time-varying channels. One conceivable reason may be that the self-interference (SI) problem in IBFD-relied systems is still intractable. This becomes even more challenging in the high-mobility communication scenarios. Note that, although channel aging is not considered, in \cite{mirza2018performance}, the channel acquisition relying on a hybrid IBFD and TDD mode was demonstrated to be efficient. With this hybrid scheme, partial DL transmissions are activated during UL training, leading to  an increased SE. Moreover, the effect of SI cancellation (SIC) on the performance of channel estimation was studied with the IBFD systems, showing that if SIC is insufficient, IBFD is unable to outperform the conventional TDD. Similar observations were also obtained in \cite{rajashekar2019multicarrier}. 

To overcome the weakness of both IBFD and TDD over fast time-varying communication channels, inspired by \cite{yang2009multicarrier}, we propose the multicarrier division duplex (MDD) associated with the dedicated frame design so as to relieve the channel aging problem, which constitutes the main motivation of this paper. The rationale can be briefly explained as follows. On the one hand, in MDD systems, both DL and UL transmissions can occur concurrently within the same frequency band but on different subcarriers. Hence, when needed, UL pilots can be continuously (or frequently) transmitted during DL transmissions. Therefore, CSI can be updated in time and does not become outdated as in TDD systems. On the other hand, MDD is capable of circumventing the stringent requirement for SIC in IBFD systems. According to \cite{bharadia2013full}, in practical IBFD systems, in addition to the SIC in the propagation- and analog-domain, the digital-domain SIC has to cancel the main SI signal component by at least 30 dB. Achieving this is very power-consuming and technically demanding, especially for the relatively small-sized mobile terminals (MTs), e.g., UAVs and smartphones. By contrast,  in the MDD-assisted systems, this amount of SIC can be attained at nearly no extra cost of system resources, owing to the embedded fast Fourier transform (FFT) operation \cite{rajashekar2019multicarrier}, which allows to ideally separate UL signals from DL signals in the digital-domain. For more detailed comparison between MDD and the existing duplex modes, e.g., TDD, FDD and IBFD, in terms of system architecture, initial access of physical layer and resource-allocation, the readers are referred to our previous works~\cite{rajashekar2019multicarrier,li2020self,li2021resource,li2021multicarrier}.

Owing to the aforementioned merits of MDD, it is important to investigate the MDD's implementation in fast time-varying scenarios, and to unlock its advantages over the conventional TDD and IBFD systems. Therefore, our main contributions of this paper are summarized as follows:
\begin{itemize}

\item {\em Firstly}, a model for the MDD-based multiuser mMIMO systems communicating over time-varying channels is presented. To alleviate the channel aging problem, we propose two types of frame structures dedicated for MDD, which enable UL pilots (or UL data transmissions) to occur concurrently with DL transmissions, so that CSI can be promptly updated whenever needed. For comparison, the corresponding TDD frame structures related to the 3GPP standards \cite{etsi2013136,3gpp2017nr} as well as their modified forms for supporting different mobility scenarios are introduced.

\item {\em Secondly}, to operate with the proposed frame structures, we introduce two finite impulse response WPs that consider both channel estimation and residual SI errors, namely the general WP and the decision-directed WP (DD-WP), dedicated to MDD systems. To be more specific, the general WP directly predicts the time-domain CSI based on the observations collected from the UL channel estimation. By contrast, the DD-WP leverages both the UL pilots and the detected UL data symbols to predict the frequency-domain CSI. Along with these WPs, the impact of residual SI and the order of WPs on the performance of MDD systems are studied and compared. 

\item {\em Thirdly}, by assuming the zero-forcing (ZF) precoding for DL transmission and the maximum ratio combining (MRC) for UL detection, the closed-form expressions for approximating the lower bounded average sum rates of both the TDD and MDD systems are derived, when the proposed WPs are operated with two general types of frame structures. 

\end{itemize}

Our studies and simulation results show that TDD systems suffer from the channel aging problem, whose performance degrades significantly with time, when channels vary fast. By contrast, the MDD systems endowed with the FD capability can effectively mitigate the channel aging problem and hence, are capable of significantly outperforming their TDD counterparts, when communicating over fast time-varying channels. Moreover, the studies demonstrate that the SIC in FD systems plays a paramount role in channel prediction. In the case of imperfect SIC, MDD becomes more competent than IBFD for operation in high-mobility communications scenarios.

The rest of the paper is organized as follows. In Section~\ref{section-MDDCEF-sys}, the model for the MDD-based system is described. In Section~\ref{sec:MDD-CEF:Frame}, two general frame structures for TDD and MDD systems are introduced. Section~\ref{sec:MDD-CEF:CSIAcqui} presents the principles of channel estimation and two approaches for channel prediction. Section~\ref{sec:MDD-CEF:AER} analyzes the lower bounded average sum rates of MDD and TDD systems. Section~\ref{sec:MDD-CEF:Sim} provides the performance results to compare TDD, MDD and IBFD systems. Finally, conclusions are drawn in Section~\ref{sec:MDD-CEF:con}.

Throughout the paper, the following notations are used: $\pmb{A}$,
$\pmb{a}$ and $a$ stand for matrix, vector, and scalar, respectively;
$\mathcal{A}$ and $\left|\mathcal{A}\right|$ represent the set and the cardinality of set, respectively; $(\pmb{a})_{i}$ denotes the $i$-th element of $\pmb{a}$; $\pmb{A}^{(i,:)}$, $\pmb{A}^{(:,j)}$ and $(\pmb{A})_{i,j}$ denote the $i$-th row, the $j$-th column and the $(i,j)$-th element of $\pmb{A}$, respectively; Furthermore, 
$\left|\pmb{A}\right|$, $\pmb{A}^*$, $\pmb{A}^T$,
$\pmb{A}^{-1}$ and $\pmb{A}^H$ represent, respectively, the
determinant, complex conjugate, transpose, inverse and Hermitian transpose of $\pmb{A}$; The Euclidean norm of a vector and the Frobenius norm of a matrix are denoted as $\left\|\cdot\right\|_2$ and $\left\|\cdot\right\|_F$, respectively; $\pmb{I}_{N}$ denotes a $(N\times N)$ identity matrix;
$\mathcal{CN}(\pmb{0},\pmb{A})$ represents the zero-mean complex
Gaussian distribution with covariance matrix $\pmb{A}$; Furthermore,
$\text{Tr}(\cdot)$, $\text{log}(\cdot)$ and $\mathbb{E}[\cdot]$ denote
the trace, logarithmic and expectation operators, respectively.

\section{System Model}\label{section-MDDCEF-sys}

Consider a single-cell mMIMO OFDM system having a base station (BS) equipped with $N$ antennas and $D$ single-antenna MTs randomly distributed in the cell. The system is operated in the MDD mode, allowing DL and UL to communicate concurrently in the same frequency band but on different subcarriers. Following the concept of mMIMO, the number of antennas at BS is assumed to be much larger than the number of served MTs \cite{sanguinetti2019toward}, i.e., $N \gg D$. Furthermore, based on the principles of MDD~\cite{rajashekar2019multicarrier}, subcarriers are divided into two mutually exclusive subsets, namely a DL subcarrier subset $\mathcal{M}$ with $M$ subcarriers, and a UL subcarrier subset $\bar{\mathcal{M}}$ with $\bar{M}$ subcarriers, i.e., $|\mathcal{M}|=M$ and $|\bar{\mathcal{M}}|=\bar{M}$. The total number of subcarriers is expressed as $M_{\text{sum}}=M+\bar{M}$. Similar to the other works on resource-allocation, e.g.,\cite{nam2015joint,li2016energy,zhang2004multiuser}, the DL/UL subcarrier allocation (SA) is carried out at BS based on the CSI and the requirements of MTs' quality of services. The allocation results are informed to MTs through control channels during the initial access procedure. Below we assume that the DL/UL SA results have been obtained, and that each MT knows its assigned subcarriers. The readers interested in the SA in MDD systems are referred to \cite{li2021resource} for details.

\subsection{Channel Model}

We assume that the channels are frequency-selective in terms of the $M_{\text{sum}}$ subcarriers but each subcarrier experiences flat fading. To consider the influence of channel aging, we assume that the channel coefficients do not change within one OFDM symbol, but vary from one symbol to the next. Therefore, the time-domain CSI of the $L$-tap channel between the $d$-th MT and the $n$-th BS antenna over the $i$-th OFDM symbol duration can be expressed as 
\begin{equation}\label{eq:MDD-CEF-TDCIR}
\pmb{g}_{n,d}[i]=\left[g_{n,d}[i,1],...,g_{n,d}[i,l],...,g_{n,d}[i,L]\right]^T
\end{equation}
where $g_{n,d}[i,l]=\nu_{n,d}[i,l]\sqrt{\beta_{d}}$ and $\nu_{n,d}[i,l]\sim \mathcal{CN}(0,1/L)$ is the small-scale fading, while $\beta_{d}$ represents the large-scale fading, which only depends on the distance between MT $d$ and BS and is assumed to remain constant over one communication frame. For any user-antenna pair, the channels of different taps are assumed to be independent. Hence we have $\pmb{R}_{g}^d=\mathbb{E}\left[\pmb{g}_{n,d}[i]\pmb{g}_{n,d}^H[i]\right]=\frac{\beta_{d}}{L}\pmb{I}_L$.

According to the principles of OFDM~\cite{yang2009multicarrier}, the
frequency-domain CSI $\pmb{h}_{n,d}[i]$ over the $i$-th OFDM symbol duration can be obtained as
\begin{equation}
\label{eq:MDD-CEF-FDCIR}
\pmb{h}_{n,d}[i]=\pmb{F}\pmb{\varPsi}\pmb{g}_{n,d}[i]
\end{equation}
where $\pmb{F} \in \mathbb{C}^{M_{\text{sum}}\times M_{\text{sum}}}$
is the FFT matrix with $\left(\pmb{F}\right)_{p,q}=\frac{1}{\sqrt{M_{\text{sum}}}}e^{-j2\pi(p-1)(q-1)/M_{\text{sum}}}$, $\pmb{\varPsi}\in \mathbb{C}^{{M_{\text{sum}}\times
    L}}$ is constructed by the first $L$ columns of
$\pmb{I}_{M_\text{sum}}$. Furthermore, the DL subchannels
$\pmb{h}_{n,d}^{\text{DL}}[i]$ and UL subchannels
$\pmb{h}_{n,d}^{\text{UL}}[i]$ between the $n$-th antenna at BS and the
$d$-th MT over the $i$-th OFDM symbol can be obtained from (\ref{eq:MDD-CEF-FDCIR}), which can be expressed as
\begin{align}\label{eq:MDD-CEF-FDCIR-DL}
\pmb{h}_{n,d}^{\text{DL}}[i]&={\pmb{\Phi}_{\text{DL}}}\pmb{h}_{n,d}[i] \nonumber \\
&=\left[h_{n,d}[i,1],...,h_{n,d}[i,m],...,h_{n,d}[i,M]\right]^T\\
\label{eq:MDD-CEF-FDCIR-UL}
\pmb{h}_{n,d}^{\text{UL}}[i]&={\pmb{\Phi}_{\text{UL}}}\pmb{h}_{n,d}[i] \nonumber \\
&=\left[h_{n,d}[i,1],...,h_{n,d}[i,\bar{m}],...,h_{n,d}[i,\bar{M}]\right]^T
\end{align} 
where $\pmb{\Phi}_{\text{DL}}=\pmb{I}_{M_\text{sum}}^{(\mathcal{M},:)}$ and $\pmb{\Phi}_{\text{UL}}=\pmb{I}_{M_\text{sum}}^{(\mathcal{\bar{M}},:)}$ are the mapping matrices,
constructed from $\pmb{I}_{M_\text{sum}}$ by choosing its rows corresponding to the particular subcarriers assigned to DL and UL, respectively.

Since in the analog-domain the MDD systems are operated in the FD mode, there is SI at both BS and MTs. In particular, although the MDD-based systems employ the inherent advantage to suppress SI in the digital-domain by the FFT operation\footnote{In IBFD-based systems, in order to mitigate the digital-domain SI, the receiver has to estimate the channel between the DAC at transmitter and the ADC at receiver so as to reconstruct the transmitted signal, which is then subtracted from the received signal \cite{sabharwal2014band}. Explicitly, this process imposes a heavy burden on system overhead, especially when a system with a large number of subcarriers is considered.}, the DL and UL subcarrier signals are coupled in the analog-domain before the ADC at the BS receiver. As in the traditional IBFD systems \cite{kolodziej2019band}, without sufficient SIC in the propagation- and analog-domain prior to ADC, the high-power SI imposed by the DL subcarriers from the BS transmitter may overwhelm the desired UL signals from MTs, which leads to the inefficient operation of ADC and consequently significant quantization error in digital-domain. Due to the short separation between transmitter and receiver, the SI channels are assumed to be flat fading and identical to all subcarriers, and also quasi-static within one frame. Specifically, the SI channels at BS and MTs follow the distributions of $(\pmb{H}_{\text{SI}})_{i,j}\sim \mathcal{CN}(0,1)$, where $\pmb{H}_{\text{SI}} \in \mathbb{C}^{N\times N} $, and $h_{\text{SI}}\sim \mathcal{CN}(0,1)$, respectively.  

\subsection{Channel Aging}

As we stated in Section \ref{section-MDDCEF-intro}, the relative mobility between BS and MTs leads to time-varying channels, causing that the channel varies between the time when it is estimated and the time when the estimated CSI is applied for transmissions. This is the well-known channel aging problem. To model the channel aging, we introduce the Jakes model, which has the normalized discrete-time autocorrelation function at the BS given by \cite{yuan2020machine}
\begin{equation}
R[k]=J_0(2\pi f_DT_s|k|),
\end{equation}
where $J_0(\cdot)$ is the zeroth-order Bessel function of the first kind, $T_s$ is the OFDM symbol duration, $f_D$ is the maximum Doppler frequency shift, and $|k|$ is the delay in terms of the number of symbols.

Furthermore, for the sake of analyzing MDD frame structures with WP and DD-WP in the following sections, we adopt the autoregressive model of order 1, denoted as AR(1), for approximating the temporal correlation between adjacent symbols. In this case, the channel coefficient $\pmb{g}_{n,d}[i]$ of \eqref{eq:MDD-CEF-TDCIR} can be formulated as \cite{baddour2005autoregressive} 
\begin{equation}\label{eq:MDD-CEF:AR}
\pmb{g}_{n,d}[i]=\alpha \pmb{g}_{n,d}[i-1]+\pmb{v}_{n,d}[i]
\end{equation}
where $\alpha=J_0(2\pi f_DT_s)$, which is assumed to be perfectly known at BS, $\pmb{v}_{n,d}[i]$ is an uncorrelated channel error vector due to channel aging, which can be modeled as a stationary Gaussian random process with i.i.d. entries and the distribution of $\mathcal{CN}\left(\pmb{0}, (1-\alpha^2)\pmb{R}_{g}^d\right)$. According to the channel model of \eqref{eq:MDD-CEF:AR}, $\pmb{g}_{n,d}[i]$ is a stationary Markov random process, and has the auto-correlation of $\mathbb{E}\left[\pmb{g}_{n,d}[i-q]\pmb{g}_{n,d}^H[i-p]\right]=\alpha^{\left|p-q\right|}\pmb{R}_{g}^d$. Moreover, according to \eqref{eq:MDD-CEF-FDCIR-DL} and \eqref{eq:MDD-CEF-FDCIR-UL}, we can obtain the DL subchannels satisfying 
\begin{equation}\label{eq:MDD-CEF:hndik}
h_{n,d}[i,m]=\alpha h_{n,d}[i-1,m] + \underbrace{\pmb{\psi}_m\pmb{v}_{n,d}[i]}_{\tilde{v}_{n,d}^m[i]}
\end{equation} 
where $\pmb{\psi}_m=\pmb{I}_{\text{sum}}^{(m,:)}\pmb{F}\pmb{\varPsi}$ and $\tilde{v}_{n,d}^m[i] \in \mathcal{CN}\left(0, (1-\alpha^2)R_{h}^d\right)$ with $R_{h}^d=\mathbb{E}\left[h_{n,d}[i,m]h_{n,d}^H[i,m]\right]={\beta_d}/{M_{\text{sum}}}$. Similarly, the UL subchannels $h_{n,d}[i,\bar{m}]$ can be generated.

\subsection{Transmission Model}
\subsubsection{Downlink Transmission}
For DL transmission, the signal $\pmb{s}_{\text{DL}}[i,m] \in \mathbb{C}^{N \times 1}$ transmitted by BS on the $m$-th subcarrier in the $i$-th OFDM symbol duration is 
\begin{equation}\label{eq:MDD-CEF:sDL}
\pmb{s}_{\text{DL}}[i,m]=\sqrt{p_{\text{DL}}}\pmb{F}^{\text{ZF}}[i,m]\pmb{x}[i,m]
\end{equation}
where $p_{\text{DL}}$ is the power budget per DL subcarrier and the total BS transmit power is $P_{\text{DL}}=p_{\text{DL}}M$,
$\pmb{x}[i,m]=\left[x_1[i,m],...,x_D[i,m]\right]^T$ contains the information symbols normalized to satisfy $\mathbb{E}\left[\pmb{x}[i,m]\pmb{x}[i,m]^H\right]=\pmb{I}_{D}$. We assume that the ZF precoding is applied at transmitter, giving $\pmb{F}^{\text{ZF}}[i,m]=\pmb{H}_{\text{DL}}^H[i,m]\left(\pmb{H}_{\text{DL}}[i,m]\pmb{H}_{\text{DL}}^H[i,m]\right)^{-1}$, where $\pmb{F}^{\text{ZF}}[i,m]=\left[\pmb{f}_1[i,m],...,\pmb{f}_D[i,m]\right]$ with the constraint of $\left\|\pmb{f}_d[i,m]\right\|_2=1/\sqrt{D}$, such that $\mathbb{E}\left[\left\|\pmb{s}_{\text{DL}}[i,m]\right\|_2^2\right]=p_{\text{DL}}$, and $\pmb{H}_{\text{DL}}[i,m]=\left[\pmb{h}_{1}[i,m],...,\pmb{h}_{d}[i,m],...\pmb{h}_{D}[i,m]\right]^H$, where $\pmb{h}_{d}[i,m]=\left[h_{1,d}[i,m],h_{2,d}[i,m],...,h_{N,d}[i,m]\right]^T$. It is noteworthy that the analysis in this paper is based on the equal power allocation among MTs. Intuitively, an appropriate power allocation method, e.g., that in our previous work \cite{li2021resource},  will increase the achievable sum rates, owing to both user and subcarrier diversity being exploited.
The received signal at the $d$-th MT can be expressed as 
\begin{equation}
\label{eq:MDD-CEF:ydDetect}
\begin{split}
&y_d[i,m]= \sqrt{p_{\text{DL}}}\pmb{h}_{d}^H [i,m]\pmb{f}_d[i,m]x_d[i,m] \\
&+\sqrt{p_{\text{DL}}}\sum_{k\neq d}\pmb{h}_{d}^H[i,m]\pmb{f}_k[i,m]x_k[i,m]+z^{\text{SI}}_d[i]+z_d[i,m]
\end{split}
\end{equation}
where $z_d^{\text{SI}}[i]\sim \mathcal{CN}\left( 0,\xi_{\text{MT}}p_{\text{UL}}\bar{M}\right)$ denotes the residual SI\footnote{The residual SI consists of the combined effect of the additive noise introduced by automatic gain control (AGC), non-linearity of ADC and the phase noise generated by oscillator due to RF imperfection \cite{day2012full2}.} signal imposed by the UL transmission of MT $d$ with $p_{\text{UL}}$ being the power budget per UL subcarrier, such that the total MT transmit power is $P_{\text{UL}}=p_{\text{UL}}\bar{M}$. Note that $\xi_{\text{MT}}$ in \eqref{eq:MDD-CEF:ydDetect} and $\xi_{\text{BS}}$ in \eqref{eq:MDD-CEF:sUL} denote the SIC capability that is provided by the existing SIC techniques, such as, the antenna circulator \cite{bharadia2013full}, spatial beamforming \cite{li2020self}, dual-port polarized antenna \cite{debaillie2014analog}, multi-tap RF canceller \cite{kolodziej2016multitap}, etc., implemented in propagation- and analog-domain, as well as the FFT operation operated in digital-domain. In \eqref{eq:MDD-CEF:ydDetect}, the covariance of $z_d^{\text{SI}}[i]$ is calculated in detail in Appendix \ref{Appen:MDD:CEF-1}.  In addition, $z_d[i,m]\sim \mathcal{CN}(0,\sigma^2)$ denotes the additive white Gaussian noise. Note also that, the interference caused by the other MTs' UL transmission on the DL reception is ignored. The reason is as follows. First, in analogy to the conventional OFDMA systems \cite{nam2015joint}, the inter-MT interference (IMI) and the desired DL signal are located on the mutually orthogonal subcarriers. Secondly, the large-scale fading of MT-MT link and the MT's relatively lower transmit power allow the received signal of an MT to efficiently pass the ADC. Consequently, the IMI can be readily removed by the FFT operation in the digital-domain, when all MTs and the BS are assumed to be synchronized within an allowable time window \cite{li2021multicarrier}.

\subsubsection{Uplink Transmission}
For the UL transmission, the signal received at BS over the $\bar{m}$-th UL subcarrier and $i$-th OFDM symbol duration can be expressed as
\begin{equation}
\label{eq:MDD-CEF:sUL}
\pmb{s}_{\text{UL}}[i,\bar{m}]=\sum_{d=1}^D\pmb{h}_d[i,\bar{m}]\sqrt{p_{\text{UL}}} x_d[i,\bar{m}] + \pmb{z}^{\text{SI}}[i] + \pmb{z}[i,\bar{m}]
\end{equation}
where $\pmb{h}_{d}[i,\bar{m}]=\left[h_{1,d}[i,\bar{m}],h_{2,d}[i,\bar{m}],...,h_{N,d}[i,\bar{m}]\right]^T$, $\pmb{z}[i,\bar{m}]\sim \mathcal{CN}(\pmb{0},\sigma^2\pmb{I}_N)$. $\pmb{z}^{\text{SI}}[i]$ is the SI generated by BS transmissions, which is modeled as $\pmb{z}^{\text{SI}}[i]\sim \mathcal{CN}\left(\pmb{0},\xi_{\text{BS}}p_{\text{DL}}M\pmb{I}_N\right)$. The covariance of $\pmb{z}^{\text{SI}}[i]$ is derived in detail in Appendix \ref{Appen:MDD:CEF-1}. In this paper, we assume that the MRC is used for signal detection. Then we have  $\pmb{W}^{\text{MRC}}[i,\bar{m}]=\pmb{H}_{\text{UL}}[i,\bar{m}]$, where $\pmb{H}_{\text{UL}}[i,\bar{m}]=\left[\pmb{h}_{1}[i,\bar{m}],...,\pmb{h}_{D}[i,\bar{m}]\right]$. Let us write $\pmb{W}^{\text{MRC}}[i,\bar{m}]=\left[\pmb{w}_1[i,\bar{m}],...,\pmb{w}_D[i,\bar{m}]\right]$. Then the decision variable for MT $d$ can be expressed as
\begin{equation}
\label{eq:MDD-CEF:ydULDetect}
\begin{split}
y_d[i,\bar{m}]&=\sqrt{p_{\text{UL}}}\pmb{w}_d^H[i,\bar{m}]\pmb{h}_d[i,\bar{m}]x_d[i,\bar{m}]\\
&+\sqrt{p_{\text{UL}}}\sum_{k\neq d}\pmb{w}_d^H[i,\bar{m}]\pmb{h}_k[i,\bar{m}]x_k[i,\bar{m}] \\
&+\pmb{w}_d^H[i,\bar{m}]\pmb{z}^{\text{SI}}[i] + \pmb{w}_d^H[i,m]\pmb{z}[i,\bar{m}]
\end{split}
\end{equation}
where the second term at the right-hand side is multiuser interference, while the third term is due to SI.

\section{Frame Structure Design}\label{sec:MDD-CEF:Frame}

Due to the effect of channel aging, channel coefficients may vary from symbol to symbol, leading to poor system performance in high-mobility communication scenarios \cite{truong2013effects}. To overcome the time-varying fading problem, in conventional TDD mode, channel prediction is usually implemented after the training phase to reduce CSI error. However, when the relative velocity between transmitter and receiver increases, the channel becomes less correlated, increasing the prediction error, which in turn renders the later detected symbols less reliable. One way to mitigate this problem is to employ extra training symbols for updating the CSI during data transmission. Following this observation, if the channel aging problem becomes severe, more training symbols are inserted, which results in more frequent switching between UL training and DL data transmission, as well as the decrease of useful data rate. To this end, our main motivation of this paper is to demonstrate that the MDD mode is capable of efficiently solving the channel aging problem in fast fading communication scenarios. For demonstration and comparison, in the following subsections, we will present two types of frame structures for coping with the channel aging problem and also for analyzing the performance of MDD systems in terms of sum-rate.

\begin{figure}
\centering
\includegraphics[width=0.99\linewidth]{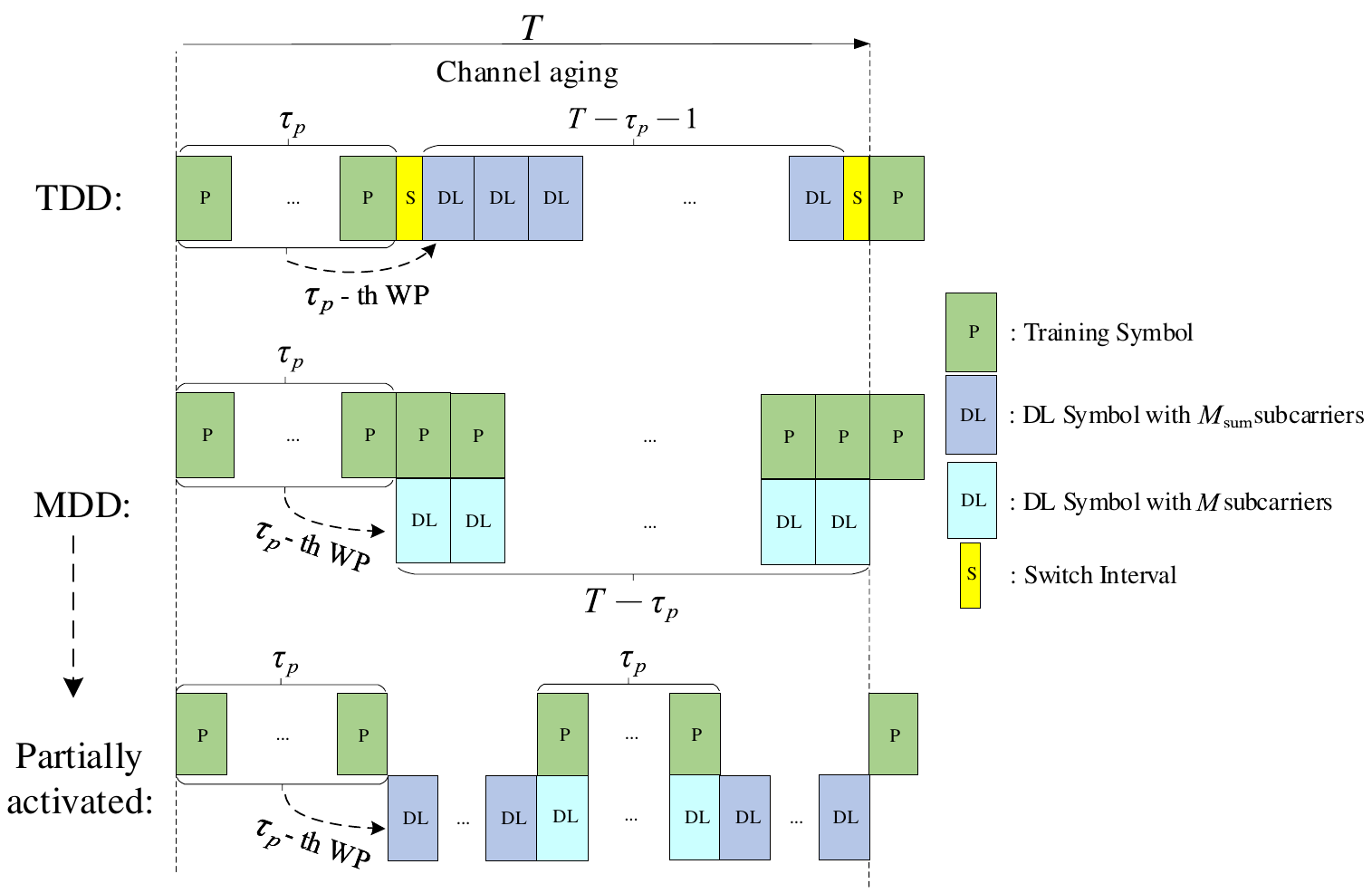} 
\caption{Type \uppercase\expandafter{\romannumeral 1} frame structure for the conventional TDD and proposed MDD systems.}
\label{figure-MDDCEF-FrameStructure_DL}
\end{figure}  


\subsection{Frame Structure: Type \uppercase\expandafter{\romannumeral 1}}
In this subsection, we consider a simple scenario, where the frame structure only includes UL training and DL transmission, as shown in Fig. \ref{figure-MDDCEF-FrameStructure_DL}. The length of one frame\footnote{Note that one frame defined in this paper may include several subframes used in 4G LTE systems or slots in 5G NR systems, where each subframe or slot contains 14 OFDM symbols. Specifically in 4G LTE, one frame consists of 10 subframes and two of them are used for training \cite{etsi2013136}. The frame structure in 5G NR is more flexible than that in 4G LTE\cite{3gpp2017nr}.} is assumed to be equal to $T$ OFDM symbol durations. 

\subsubsection{TDD} As shown in Fig. \ref{figure-MDDCEF-FrameStructure_DL}, in the TDD mode, a frame consists of $\tau_p$ training symbols transmitted by UL and $T-\tau_p-1$ symbols used for DL transmission.
The switching from UL training to DL data transmission or from DL data transmission to the next training phase incurs a fixed cost, which is assumed to be half of an OFDM symbol.\footnote{The length of the switching period in LTE can be a different value, such as 1, 2, 3, 4, 9 or 10 OFDM symbols \cite{etsi2013136}. We choose 0.5 to minimize the influence of the switching interval on the evaluation of TDD system's performance.} To cope with channel aging, the channel prediction is implemented by a $\tau_p$-th order WP\footnote{To ease analysis and facilitate comparison, only WP is considered in this paper. The Kalman predictor and the state-of-the-art deep learning approaches will be studied in our future work.} after the training phase, as shown in Fig. \ref{figure-MDDCEF-FrameStructure_DL}, and all the precoding/detection are operated at BS based on the predicted CSI. It is noteworthy that in practice, the $\tau_p$ training symbols can be first disassembled into several groups, which are then evenly distributed over one frame, so as to relieve the effect of accumulated prediction errors. However, the accompany of the increased switching intervals results in the reduced time for DL transmission and hence the decreased system's efficiency.

\subsubsection{MDD} 
In the context of MDD, as UL/DL transmissions are on different subcarriers, CSI estimation and prediction can be implemented in the FD mode. That is to say, UL pilots can be kept active to update CSI in real time within one OFDM symbol. In detail, as shown in Fig. \ref{figure-MDDCEF-FrameStructure_DL}, UL training is always activated and hence, the channels for DL transmission can be predicted by a WP based on the newest $\tau_p$ pilots. However, in low-speed scenarios, the excessive transmission of UL pilots inevitably increases energy consumption and system complexity, which is unnecessary as the channels change slowly. Moreover, in the case where only the DL transmission is needed, continuous UL training on $\bar{M}$ subcarriers leads to only $M$ of the $M_{\text{sum}}$ subcarriers being used for the DL transmission. Therefore, in practice, the pilots within one MDD frame can be partially transmitted (as illustrated by the example presented in Fig. \ref{figure-MDDCEF-FrameStructure_DL}) so that some UL subcarriers can be `borrowed' for DL data transmission to reduce the overhead of SI mitigation. In Section \ref{sec:MDD-CEF:Sim}, the TDD/MDD frame structures with different distributions of pilots will be investigated.

\begin{figure}
\centering
\includegraphics[width=0.99\linewidth]{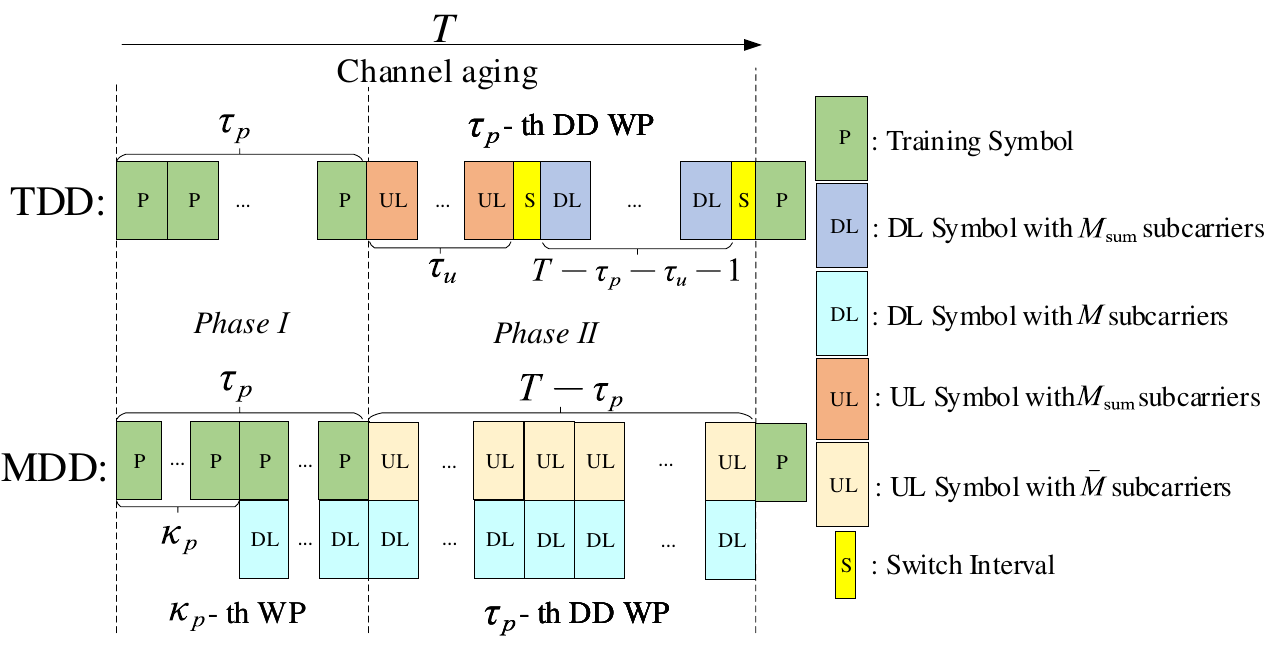} 
\caption{Type \uppercase\expandafter{\romannumeral 2} frame structure for conventional TDD and proposed MDD systems.}
\label{figure-MDDCEF-FrameStructure_DUL}
\end{figure}  

\subsection{Frame Structure: Type \uppercase\expandafter{\romannumeral 2}}

A more general frame structure consisting of the UL training followed by both UL transmission and DL transmission is considered for both TDD and MDD, as shown in Fig. \ref{figure-MDDCEF-FrameStructure_DUL}. 

\subsubsection{TDD} In the TDD mode, as shown in Fig. \ref{figure-MDDCEF-FrameStructure_DUL}, $\tau_p$ pilot symbols are transmitted in Phase \uppercase\expandafter{\romannumeral 1} for initializing the transmission. During Phase \uppercase\expandafter{\romannumeral 2}, the detected UL data symbols and possibly a part of pilots are used to predict channels by a $\tau_p$-th order decision-directed (DD)-WP. With this setting, during the UL transmission, the channel knowledge can be continuously updated with the DD prediction principle. By contrast, during the DL transmission, channel updating is unavailable as there is no UL transmission. Consequently, the transmissions of the $T-\tau_p-\tau_u-1$ DL symbols are based on the outdated CSI, which is similar to the situation with the Type \uppercase\expandafter{\romannumeral 1} frame structure.

\subsubsection{MDD} As shown in Fig. \ref{figure-MDDCEF-FrameStructure_DUL}, the Type \uppercase\expandafter{\romannumeral 2} frame structure in MDD systems can be designed in a very efficient way. Firstly, for fair comparison with TDD, in Phase \uppercase\expandafter{\romannumeral 1}, $\tau_p$ pilot symbols are still transmitted for the implementation of a $\tau_p$-th order DD-WP in Phase \uppercase\expandafter{\romannumeral 2}. However, different from TDD, during the transmission of pilot symbols, DL symbols can also be transmitted with the aid of the CSI predicted by a $\kappa_p$-th order WP, where $1\leq \kappa_p \leq \tau_p$. In the extreme case, when $\kappa_p=1$, $\tau_p-1$ more DL symbols than TDD can be transmitted in Phase \uppercase\expandafter{\romannumeral 1}, which may significantly increase the SE. During Phase \uppercase\expandafter{\romannumeral 2}, both DL and UL can transmit $T-\tau_p$ symbols. Furthermore, a continuously updated $\tau_p$-th order DD-WP can be implemented to predict the channels using the pilots and the detected UL symbols.

\section{Channel Estimation and Prediction In MDD systems}\label{sec:MDD-CEF:CSIAcqui}
In this section, we study the channel estimation and prediction in MDD systems, when the proposed frame structures in Section \ref{sec:MDD-CEF:Frame} are introduced.

\subsection{Channel Estimation}
Let us assume that all MTs synchronously transmit their frequency-domain pilot sequences (FDPS) over $\bar{M}$ UL subcarriers. Specifically, the FDPS transmitted by the $d$-th MT is expressed as $\pmb{x}_{p,d}=\left[x_{p,d}[1],...,x_{p,d}[\bar{m}],...,x_{p,d}[\bar{M}]\right]^T$. Then, the received signal at the $n$-th antenna of BS corresponding to the $i$-th symbol is given by
\begin{equation}\label{eq:MDD-CEF:ypnCE}
\begin{split}
\pmb{y}_{p,n}[i]&=\sqrt{p_{\text{UL}}}\sum\limits_{d=1}^D\pmb{X}_{p,d}\pmb{h}_{n,d}^{\text{UL}}[i]+\pmb{z}^{\text{SI}}_n[i]+\pmb{z}_{p,n}[i] \\
&= \sqrt{p_{\text{UL}}}\sum\limits_{d=1}^D\pmb{X}_{p,d}{\pmb{\Phi}_{\text{UL}}}\pmb{F}\pmb{\varPsi}\pmb{g}_{n,d}[i]+\pmb{z}^{\text{SI}}_n[i]+\pmb{z}_{p,n}[i]
\end{split}
\end{equation}
where $\pmb{X}_{p,d}=\text{diag}\left\{\pmb{x}_{p,d}\right\}$,  
 $\pmb{z}_{p,n}[i]\sim \mathcal{CN}(\pmb{0},\sigma^2\pmb{I}_{\bar{M}})$. Note that, in MDD systems, BS may experience SI during UL training, if DL symbols are transmitted at the same time. For example, in the frame structure as shown in Fig. \ref{figure-MDDCEF-FrameStructure_DL}, when $1 \leq i \leq \tau_p$, we have $\pmb{z}^{\text{SI}}_n[i]=0$, as there is no DL transmissions. By contrast, when $\tau_p<i\leq T$, $\pmb{z}^{\text{SI}}_n[i] \sim \mathcal{CN}\left( 0,\xi_{\text{BS}}p_{\text{DL}}M\pmb{I}_{\bar{M}}\right)$.

According to our previous study \cite{li2020self}, the FDPS of MT $d$ can be designed as
\begin{equation}
\pmb{x}_{p,d}=\left[1,e^{j2\pi\frac{(d-1)\xi}{\bar{M}}},e^{j2\pi\frac{2(d-1)\xi}{\bar{M}}},\cdots, e^{j2\pi\frac{(\bar{M}-1)(d-1)\xi}{\bar{M}}}\right]^T
\end{equation}
where $\xi=\left\lfloor \frac{\bar{M}}{D}\right\rfloor$. With the assumption that $\bar{M}\geq DL$ and the $\bar{M}$ number of UL subcarriers are evenly distributed, $\pmb{P}_d=\pmb{X}_{p,d}{\pmb{\Phi}_{\text{UL}}}\pmb{F}\pmb{\varPsi}$ for $d=1,...,D$ are mutually orthogonal, i.e., 
\begin{align}
\begin{cases}
\pmb{P}_d^H \pmb{P}_d =\frac{\bar{M}}{M_{\text{sum}}} \pmb{I}_L,  \\
\pmb{P}_d^H \pmb{P}_k= \pmb{0}_L,~\forall~ {d}\neq{k}
\end{cases}
\end{align}
Then, the noisy observation of $\pmb{g}_{n,d}[i]$ is given by
\begin{align}\label{eq:MDD-CEF:gndobser}
\tilde{\pmb{y}}_{p,n}^d[i]&=\pmb{P}_d^H \pmb{y}_{p,n}[i] \nonumber \\
&=\frac{\sqrt{p_{\text{UL}}}\bar{M}}{M_{\text{sum}}}\pmb{g}_{n,d}[i]+\pmb{P}_d^H (\pmb{z}^{\text{SI}}[i]+\pmb{z}_{p,n}[i])
\end{align} 
Correspondingly, the MMSE estimate to $\pmb{g}_{n,d}[i]$ is given by
\begin{equation}
\hat{\pmb{g}}_{n,d}[i]=\frac{\frac{\sqrt{p_{\text{UL}}}\bar{M}}{M_{\text{sum}}}\frac{\beta_d}{L}}{\frac{p_{\text{UL}}\bar{M}^2}{M_{\text{sum}}^2}\frac{\beta_d}{L}+\frac{\xi_{\text{BS}}p_{\text{DL}}M\bar{M}}{M_{\text{sum}}}+\frac{\bar{M}}{M_{\text{sum}}}\sigma^2}\tilde{\pmb{y}}_{p,n}^d[i]
\end{equation}



\subsection{Channel Prediction}\label{sec:MDD-CEF:WP}
\subsubsection{General Wiener Predictor}
In this subsection, we focus on the prediction of $\pmb{g}_{n,d}[i+1]$ based on the current and previous received training signals, which are expressed as $\bar{\pmb{y}}_{p,n}^d[i]=\left[\tilde{\pmb{y}}_{p,n}^{d,H}[i], \tilde{\pmb{y}}_{p,n}^{d,H}[i-1],...,\tilde{\pmb{y}}_{p,n}^{d,H}[i+1-\tau_p]\right]^H$, where $\tilde{\pmb{y}}_{p,n}^{d}[i]$ is given by \eqref{eq:MDD-CEF:gndobser}. We first assume a $\tau_p$-th order WP, with the weights $\pmb{V}_{n,d}=\left[\pmb{V}_{n,d}^1,...,\pmb{V}_{n,d}^q,...,\pmb{V}_{n,d}^{\tau_p}\right]$, $\pmb{V}_{n,d}^q \in \mathbb{C}^{L\times L}$. Then, the predictor can be formulated as
\begin{equation}
\check{\pmb{g}}_{n,d}^{\text{WP}}[i+1]=\sum\limits_{q=1}^{\tau_p}\pmb{V}_{n,d}^q\tilde{\pmb{y}}_{p,n}^{d}[i+1-q]=\pmb{V}_{n,d}\bar{\pmb{y}}_{p,n}^d[i]
\end{equation}

According to the principles of Wiener filter \cite{truong2013effects,haykin2005adaptive}, the solution to $\pmb{V}_{n,d}$ is 
\begin{equation}\label{eq:MDD-CEF-Vnd}
\pmb{V}_{n,d}=\pmb{R}_{g\bar{y}}^d[1]\left(\pmb{R}_{\bar{y},i}^d[0]\right)^{-1}
\end{equation}
where the cross-correlation between the real channel of the next symbol and the observations is 
\begin{equation}\label{eq:MDD-CEF:Rgyd1}
\pmb{R}_{g\bar{y}}^d[1]=\mathbb{E}\left[\pmb{g}_{n,d}[i+1]\bar{\pmb{y}}_{p,n}^{d,H}[i]\right]=\frac{\sqrt{p_{\text{UL}}}\bar{M}}{M_{\text{sum}}}[\pmb{\delta}(\tau_p,\alpha)\otimes \pmb{R}_g^d]
\end{equation}
where $\pmb{\delta}(\tau_p,\alpha)=[\alpha,\alpha^2,...,\alpha^{\tau_p}]$. In \eqref{eq:MDD-CEF-Vnd}, the autocorrelation matrix of the training signals is 
\begin{align}
\pmb{R}_{\bar{y},i}^d[0]&=\bar{\pmb{y}}_{p,n}^{d}[i]\bar{\pmb{y}}_{p,n}^{d,H}[i] \nonumber \\
&=\pmb{\xi}(\tau_p,\alpha)\otimes (\frac{p_{\text{UL}}\bar{M}^2}{M_{\text{sum}}^2}\pmb{R}_g^d)+\frac{\bar{M}\sigma^2}{M_{\text{sum}}}\pmb{I}_{L\tau_p}+\pmb{\Xi}_{\text{SI}}[i]
\end{align}
where 
\begin{align}
\pmb{\xi}(\tau_p,\alpha)&=
\left[
\begin{array}{cccc}
1 & \alpha & \cdots & \alpha^{\tau_p-1} \\
\alpha & 1 & \cdots & \alpha^{\tau_p-2} \\
\vdots & \vdots & \ddots & \vdots \\
\alpha^{\tau_p-1} & \alpha^{\tau_p-2} & \cdots & 1
\end{array}\right]
\end{align}
It can be observed that in the Type \uppercase\expandafter{\romannumeral 1} frame structure for fast-fading channels, the amount of SI is dependent on the observation time, which leads to 
\begin{align}
\pmb{\Xi}_{\text{SI}}[i]&=
\begin{cases}
\pmb{0}_{L\tau_p \times L\tau_p}, \ \ i=\tau_p \\
\left[\begin{array}{cc}
\frac{\xi_{\text{BS}}p_{\text{DL}}M\bar{M}}{M_{\text{sum}}}\pmb{I}_{(i-\tau_p)L}&\pmb{0} \\
\pmb{0} & \pmb{0}\\
\end{array}
\right], \ \ \tau_p<i\leq 2\tau_p \\
\frac{\xi_{\text{BS}}p_{\text{DL}}M\bar{M}}{M_{\text{sum}}}\pmb{I}_{\tau_pL}, \ \ i>2\tau_p \label{eq:MDD-CEF:XiSI}
\end{cases}
\end{align}

According to the properties of WP, $\pmb{g}_{n,d}[i+1]$ can be orthogonally decomposed into
\begin{equation}\label{eq:MDD-CEF:gWPOrth}
\pmb{g}_{n,d}[i+1] = \check{\pmb{g}}_{n,d}^{\text{WP}}[i+1]+ \check{\pmb{v}}_{n,d}[i+1]
\end{equation}
where $\check{\pmb{v}}_{n,d}[i+1] \sim \mathcal{CN}(\pmb{0},\pmb{R}_g^d-\pmb{\Upsilon}_{d,i})$ is the uncorrelated channel prediction error vector with $\pmb{\Upsilon}_{d,i}=\pmb{R}_{g\bar{y}}^d[1]{\pmb{R}_{\bar{y},i}^d}^{-1}[0]\pmb{R}_{g\bar{y}}^{d,H}[1]$. 

Similar to \eqref{eq:MDD-CEF:hndik}, the predicted DL subchannels can be attained, which are expressed as
\begin{equation}\label{eq:MDD-CEF:hWPOrth}
h_{n,d}[i+1,m] = \check{h}_{n,d}^{\text{WP}}[i+1,m]+ \underbrace{\pmb{\psi}_m\check{\pmb{v}}_{n,d}[i+1]}_{\check{v}^{\text{WP}}_{n,d}[i+1,m]}, \ m=1,...,M
\end{equation}
where $\check{h}_{n,d}^{\text{WP}}[i+1,m]\sim \mathcal{CN}\left(0,\sigma^2_{\check{h}_{d,i,m}^{\text{WP}}}\right)$ with $\sigma^2_{\check{h}_{d,i,m}^{\text{WP}}}=\pmb{\psi}_m\pmb{\Upsilon}_{d,i}\pmb{\psi}_m^H$, and $\check{v}^{\text{WP}}_{n,d}[i+1,m] \sim \mathcal{CN}(0,\sigma^2_{\check{v}_{d,i,m}^{\text{WP}}})$ with $\sigma^2_{\check{v}_{d,i,m}^{\text{WP}}}=R_h^d-\sigma^2_{\check{h}_{d,i,m}^{\text{WP}}}$. Moreover, as the channels of different antennas are assumed to be independent, the channels between BS antennas and the $d$-th MT can be collected to a vector as 
\begin{equation}\label{eq:MDD-CEF:hWPvec}
\check{\pmb{h}}^{\text{WP}}_d[i+1,m]=\pmb{h}_d[i+1,m]+\check{\pmb{v}}_{d}^{\text{WP}}[i+1,m]
\end{equation}
where $\check{\pmb{h}}^{\text{WP}}_d[i+1,m]\sim \mathcal{CN}(\pmb{0},\sigma^2_{\check{h}_{d,i,m}^{\text{WP}}}\pmb{I}_N)$ and $\check{\pmb{v}}_{d}^{\text{WP}}[i+1,m]\sim \mathcal{CN}(\pmb{0},\sigma^2_{\check{v}_{d,i,m}^{\text{WP}}}\pmb{I}_N$).

\subsubsection{Decision-Directed Wiener Predictor}
In the proposed Type \uppercase\expandafter{\romannumeral 2} frame structure, as UL transmission is activated, channels can be predicted in the decision-directed (DD) principles \cite{schafhuber2005mmse}. Assume that a $\tau_p$-th order DD-WP is performed based on the $\tau_p$ latest detected symbols received by BS. According to \eqref{eq:MDD-CEF:sUL}, the received signal by the $n$-th antenna of BS over the $\bar{m}$-th UL subcarrier and the $i$-th symbol duration is 
\begin{equation}\label{eq:MDD-CEF:snim}
s_n[i,\bar{m}]=\sqrt{p_{\text{UL}}}\pmb{x}^T[i,\bar{m}]\pmb{h}_n[i,\bar{m}]+z^{\text{SI}}[i]+z[i,\bar{m}]
\end{equation}
where $\pmb{x}[i,\bar{m}]=\left[{x}_1[i,\bar{m}],...,{x}_D[i,\bar{m}]\right]^T$, $\pmb{h}_n[i,\bar{m}]=\left[h_{n,1}[i,\bar{m}],...,h_{n,D}[i,\bar{m}]\right]^T$ and $z^{\text{SI}}[i] \sim\mathcal{CN}\left( 0,\xi_{\text{BS}}p_{\text{DL}}M\right)$.

Based on the DD signals of \eqref{eq:MDD-CEF:snim}, a $\tau_p$-th order DD-WP is employed to predict the channels $\pmb{h}_n[i+1,\bar{m}]$ as $\check{\pmb{h}}_n^{\text{DD}}[i+1,\bar{m}]=\pmb{V}_{\text{DD}}\bar{\pmb{s}}_n[i,\bar{m}]$, where $\bar{\pmb{s}}_n[i,\bar{m}]=[s_n[i,\bar{m}],...,s_n[i+1-\tau_p,\bar{m}]]^T$, and $\pmb{V}_{\text{DD}}$ can be expressed as 
\begin{equation}\label{eq:MDD-CEF:VDD}
\pmb{V}_{\text{DD}}=\pmb{R}_{h\bar{s}}[1,\bar{m}]\pmb{R}_{\bar{s},i}^{-1}[0,\bar{m}]
\end{equation}
with $\pmb{R}_{h\bar{s}}[1,\bar{m}]=\mathbb{E}\left[\pmb{h}_n[i+1,\bar{m}]\bar{\pmb{s}}_n^H[i,\bar{m}]\right]$ and $\pmb{R}_{\bar{s},i}[0,\bar{m}]=\mathbb{E}\left[\bar{\pmb{s}}_n[i,\bar{m}]\bar{\pmb{s}}_n^H[i,\bar{m}]\right]$, which are detailed below. 

First, the cross-correlation matrix in \eqref{eq:MDD-CEF:VDD} can be obtained with the aid of \eqref{eq:MDD-CEF:hndik} and \eqref{eq:MDD-CEF:snim}, which is
\begin{equation}\label{eq:MDD-CEF:Rhsbar1}
\begin{split}
&\pmb{R}_{h\bar{s}}[1,\bar{m}]=\mathbb{E}\left[\pmb{h}_n[i+1,\bar{m}]\bar{\pmb{s}}_n^H[i,\bar{m}]\right] \\
&=\sqrt{p_\text{UL}}\left[\alpha \pmb{R}_h{\pmb{x}}^*[i,\bar{m}],...,\alpha^{\tau_p} \pmb{R}_h{\pmb{x}}^*[i+1-\tau_p,\bar{m}]\right] \\
&= \sqrt{p_\text{UL}}\pmb{A}(\tau_p,\alpha) \pmb{Q}(\tau_p,D)\pmb{B}^H(\tau_p,D)
\end{split}
\end{equation} 
where 
\begin{equation}
\begin{split}
&\pmb{R}_h=\text{diag}(R_{h}^1,...,R_{h}^D), \ \pmb{Q}(\tau_p,D)=\pmb{R}_h\otimes \pmb{I}_{\tau_p} \\
&\pmb{A}(\tau_p,\alpha)=\pmb{I}_D\otimes \pmb{\delta}(\tau_p,\alpha), \\ 
&\pmb{B}(\tau_p,D)=\left[\text{diag}(\hat{\pmb{x}}_1[\bar{m}]),...,\text{diag}(\hat{\pmb{x}}_d[\bar{m}]),...,\text{diag}(\hat{\pmb{x}}_D[\bar{m}])\right]
\end{split}
\end{equation}
and in $\pmb{B}(\tau_p,D)$, $\hat{\pmb{x}}_d[\bar{m}]=\left[\hat{x}_d[i,\bar{m}],...,\hat{x}_d[i+1-\tau_p,\bar{m}]\right]^T$ are UL detected symbols. To obtain \eqref{eq:MDD-CEF:Rhsbar1}, each element in $\hat{\pmb{x}}_d[\bar{m}]$ needs to be correctly detected.  This can be achieved by measuring the reliability of each detection in terms of the likelihood value of a detected symbol. If the likelihood value is larger than a preset threshold resulting in high reliability, the symbol is then included for DD prediction. Otherwise, the symbol will not be used for DD channel prediction.

The auto-correlation matrix $\pmb{R}_{\bar{s},i}[0,\bar{m}]$ in \eqref{eq:MDD-CEF:VDD} is given by
\begin{equation}
\begin{split}
\pmb{R}_{\bar{s},i}[0,\bar{m}]&=\mathbb{E}\left[\bar{\pmb{s}}_n[i,\bar{m}]\bar{\pmb{s}}_n^H[i,\bar{m}]\right] \\
&=p_{\text{UL}}\pmb{B}(\tau_p,D)\pmb{C}(\tau_p,D)\pmb{Q}(\tau_p,D)\pmb{B}^H(\tau_p,D) \\
&+\sigma^2\pmb{I}_{\tau_p}+\pmb{\Xi}_{\text{SI}}^{\text{DD}}[i]
\end{split}
\end{equation}
where $\pmb{C}(\tau_p,D)=\pmb{I}_D \otimes \pmb{\xi}(\tau_p,\alpha)$. When the proposed Type \uppercase\expandafter{\romannumeral 2} frame structure is employed, SI is also present at the observation time, which leads to
\begin{equation}
\pmb{\Xi}_{\text{SI}}^{\text{DD}}[i]=
\begin{cases}
\left[\begin{array}{cc}
\xi_{\text{BS}}p_{\text{DL}}M\pmb{I}_{\tau_p-1}&\pmb{0} \\
\pmb{0} & 0\\
\end{array}
\right], \ \ i=\tau_p \\
\xi_{\text{BS}}p_{\text{DL}}M\pmb{I}_{\tau_p}, \ \ i>\tau_p
\end{cases}
\end{equation}

With the aid of the orthogonal decomposition, we have the expression
\begin{equation}\label{eq:MDD:CEF:hnDDOr}
\pmb{h}_n[i+1,\bar{m}]=\check{\pmb{h}}_n^{\text{DD}}[i+1,\bar{m}]+\check{\pmb{e}}^{\text{DD}}_n[i+1,\bar{m}]
\end{equation}
where $\check{\pmb{e}}^{\text{DD}}_n[i+1,\bar{m}]\sim \mathcal{CN}(\pmb{0},\pmb{R}_h-\pmb{\Theta}_i[\bar{m}])$ is the uncorrelated channel prediction error vector, and $\pmb{\Theta}_i[\bar{m}]=\pmb{R}_{h\bar{s}}[1,\bar{m}]\pmb{R}_{\bar{s},i}^{-1}[0,\bar{m}]\pmb{R}_{h\bar{s}}^H[1,\bar{m}]$, which is the covariance matrix of $\check{\pmb{h}}_n^{\text{DD}}[i+1,\bar{m}]$. Furthermore, we can write $\check{\pmb{h}}_{n,d}^{\text{UL}}[i+1]=\left[\check{h}_{n,d}^{\text{DD}}[i+1,1],...,\check{h}_{n,d}^{\text{DD}}[i+1,\bar{M}]\right]^T$, which follows the distribution of $\check{\pmb{h}}_{n,d}^{\text{UL}}[i+1]\sim \mathcal{CN}(\pmb{0},\pmb{\Gamma}_{d,i})$, with $\pmb{\Gamma}_{d,i}$ derived in detail in Appendix \ref{Appen:MDD:CEF-2}.

Having obtained the frequency-domain prediction of $\pmb{h}_{n,d}[i+1]$ over UL subcarriers, the corresponding time-domain CSI can be found as
\begin{equation}\label{eq:MDD:CEF:gndDDPre}
\check{\pmb{g}}_{n,d}^{\text{DD}}[i+1]=\pmb{J}\check{\pmb{h}}_{n,d}^{\text{UL}}[i+1]
\end{equation} 
provided that $\bar{M}\geq L$. In \eqref{eq:MDD:CEF:gndDDPre}, $\pmb{J}=\left(\left(\pmb{\Phi}_{\text{UL}}\pmb{F}\pmb{\varPsi}\right)^H\left(\pmb{\Phi}_{\text{UL}}\pmb{F}\pmb{\varPsi}\right)\right)^{-1}\left(\pmb{\Phi}_{\text{UL}}\pmb{F}\pmb{\varPsi}\right)^H$. Finally, upon following the same approach described in \eqref{eq:MDD-CEF:gWPOrth}-\eqref{eq:MDD-CEF:hWPOrth}, the DL channel vector between BS and the $d$-th MT with respect to the $m$-th DL subcarrier and the $(i+1)$ symbol duration can be predicted as $\check{\pmb{h}}^{\text{DD}}_d[i+1,m]$, which follows
\begin{equation}
\check{\pmb{h}}^{\text{DD}}_d[i+1,m]=\pmb{h}_d[i+1,m]+\check{\pmb{v}}_{d}^{\text{DD}}[i+1,m], \ m=1,...,M
\end{equation}
where $\check{\pmb{h}}_{d}^{\text{DD}}[i+1,m]\sim \mathcal{CN}(\pmb{0},\sigma^2_{\check{h}_{d,i,m}^{\text{DD}}}\pmb{I}_N)$, $\sigma^2_{\check{h}_{d,i,m}^{\text{DD}}}=\pmb{\psi}_m\pmb{J}\pmb{\Gamma}_{d,i}\pmb{J}^H\pmb{\psi}_m^H$, and $\check{\pmb{v}}_{d}^{\text{DD}}[i+1,m]\sim \mathcal{CN}(\pmb{0},\sigma^2_{\check{v}_{d,i,m}^{\text{DD}}}\pmb{I}_N$), $\sigma^2_{\check{v}_{d,i,m}^{\text{DD}}}=R_h^d-\sigma^2_{\check{h}_{d,i,m}^{\text{DD}}}$.


\section{Performance Analysis Over fast-fading channels}\label{sec:MDD-CEF:AER}
In this section, we derive the ergodic rates attainable by the MDD and TDD systems with the proposed channel acquisition, where the two types of frame structures introduced in Section \ref{sec:MDD-CEF:Frame} are respectively applied.

\subsection{Type \uppercase\expandafter{\romannumeral 1} Frame Structure}

The ergodic achievable rate of the MDD systems with the Type \uppercase\expandafter{\romannumeral 1} frame structure shown in Fig. \ref{figure-MDDCEF-FrameStructure_DL} can be expressed as 
\begin{equation}\label{eq:MDD-CEF:MDD1kp}
R^{\text{MDD-1}}(\tau_p) = \frac{1}{TM_\text{sum}}\sum\limits_{d=1}^D\sum\limits_{i=\tau_p+1}^T\sum\limits_{m=1}^M R_{d,i,m}^{\text{MDD-1}}
\end{equation}
where $R_{d,i,m}^{\text{MDD-1}} $ denotes the rate obtained by the $d$-th MT over the $m$-th DL subcarrier during the $i$-th symbol duration. Note that since no DL pilots are employed, only the expected effective channel gain is available at MTs. Hence, the mean effective channel gain is treated as the CSI for signal detection at MTs' sides. 
The effective channel gain is $\omega_{d,d}^{\text{ZF}}[i,m]=\pmb{h}_d^H[i,m]\check{\pmb{f}}_d[i,m]$, where $\check{\pmb{f}}_d[i,m]$ is derived by the ZF principle with the predicted channel $\check{\pmb{h}}^{\text{WP}}_d[i+1,m]$ obtained in \eqref{eq:MDD-CEF:hWPvec}.
Then, we can derive the lower bounded achievable ergodic rate of $R_{d,i,m}^{\text{MDD-1-LB}}$ as shown in \eqref{eq:MDD-CEF:MDDLB}.
\begin{figure*}[!t]
\hrulefill
\begin{equation}\label{eq:MDD-CEF:MDDLB}
R_{d,i,m}^{\text{MDD-1}}\geq R_{d,i,m}^{\text{MDD-1--LB}} 
=\text{log}_2\Big(1+\frac{p_{\text{DL}}\left|\mathbb{E}\left[\omega_{d,d}^{\text{ZF}}[i,m]\right]\right|^2}{p_{\text{DL}}\text{var}\left\{\omega_{d,d}^{\text{ZF}}\left[i,m\right]\right\}+p_{\text{DL}}\sum\limits_{k=1,k\neq d}^{D}\mathbb{E}\left[\left|\omega_{d,k}^{\text{ZF}}\left[i,m\right]\right|^2\right]+\text{var}\left\{z^{\text{SI}}_d\right\}+\sigma^2}\Big)
\end{equation}
\begin{align}\label{eq:MDD-CEF:MDD1LB}
R_{d,i,m}^{\text{MDD-1-LB}}\approx \text{log}_2\Big(1+\frac{p_{\text{DL}}\left(N-D+1\right)\pmb{\psi}_m\pmb{\Upsilon}_{d,i}\pmb{\psi}_m^H}{0.25p_{\text{DL}}\pmb{\psi}_m\pmb{\Upsilon}_{d,i}\pmb{\psi}_m^H+p_{\text{DL}}D\left(\frac{\beta_d}{M_{\text{sum}}}-\pmb{\psi}_m\pmb{\Upsilon}_{d,i}\pmb{\psi}_m^H\right)+\xi_{\text{MT}}p_{\text{UL}}D\bar{M}+D\sigma^2}\Big) 
\end{align}

\vspace*{-6pt}
\end{figure*}

According to Lemma 4 in \cite{khansefid2015achievable}, the expectation and variance of the effective channel gain $\omega_{d,d}^{\text{ZF}}[i,m]$ can be approximated as
\begin{align}\label{eq:MDD-CEF:Ewdd}
\mathbb{E}\left[\omega_{d,d}^{\text{ZF}}\left[i,m\right]\right]&=\mathbb{E}\left[1/(\sqrt{D}\left\|\check{\pmb{f}}_d[i,m]\right\|)\right] \nonumber\\
& \approx \sqrt{\frac{N-D+1}{D}}\sqrt{\sigma^2_{\check{h}_{d,i,m}^{\text{WP}}}}
\end{align}
and 
\begin{align}\label{eq:MDD-CEF:Varwdd}
\text{var}\left\{\omega_{d,d}^{\text{ZF}}[i,m]\right\}\approx\frac{1}{D}\Big(\frac{1}{4}\sigma^2_{\check{h}_{d,i,m}^{\text{WP}}}+\sigma^2_{\check{v}_{d,i,m}^{\text{WP}}}\Big),
\end{align}
respectively, where $\check{\pmb{f}}_d[i,m]$ is the unnormalized column of the matrix $\check{\pmb{F}}^{\text{ZF}}[i,m]$ seen in \eqref{eq:MDD-CEF:sDL}. For the interference signal from the other MTs, as seen in the denominator of \eqref{eq:MDD-CEF:MDDLB}, we have
\begin{align}\label{eq:MDD-CEF:EwddInter}
&\mathbb{E}\left[\left|\omega_{d,k}^{\text{ZF}}\left[i,m\right]\right|^2\right]\nonumber \\
&=\mathbb{E}\left[\left|\check{\pmb{h}}_d^{\text{WP},H}[i,m]\check{\pmb{f}}_k[i,m]+\check{\pmb{v}}_{d}^{\text{WP},H}[i+1,m]\check{\pmb{f}}_k[i,m]\right|^2\right] \nonumber \\
&=\mathbb{E}\left[\left|\check{\pmb{v}}_{d}^{\text{WP},H}[i+1,m]\check{\pmb{f}}_k[i,m]\right|^2\right] \nonumber \\
&=\frac{1}{D}\sigma^2_{\check{v}_{d,i,m}^{\text{WP}}}
\end{align}
Consequently, upon substituting \eqref{eq:MDD-CEF:hWPOrth}, \eqref{eq:MDD-CEF:Ewdd}, \eqref{eq:MDD-CEF:EwddInter} and \eqref{eq:MDD-CEF:Varzd} into \eqref{eq:MDD-CEF:MDDLB}, the approximation to the lower bounded achievable ergodic rate of the $d$-th MT on the $m$-th DL subcarrier within the $i$-th OFDM symbol duration can be expressed as \eqref{eq:MDD-CEF:MDD1LB}.
Finally, $R^{\text{MDD-1-LB}}(\tau_p)$ is obtained by substituting \eqref{eq:MDD-CEF:MDD1LB} into \eqref{eq:MDD-CEF:MDD1kp}.

On the other hand, the achievable ergodic rate of the TDD systems with the Type \uppercase\expandafter{\romannumeral 1} frame structure shown in Fig. \ref{figure-MDDCEF-FrameStructure_DL} can be written as
\begin{align}\label{eq:MDD-CEF:TDD1LB}
R^{\text{TDD-1}}(\tau_p) &= \frac{1}{TM_\text{sum}}\sum\limits_{d=1}^D\Big(\sum\limits_{i=\tau_p+2}^{T-1}\sum\limits_{m=1}^{M_{\text{sum}}} R_{d,i,m}^{\text{TDD-1}} \nonumber \\
&+\frac{1}{2}\sum\limits_{m=1}^{M_{\text{sum}}} \left(R_{d,\tau_p+1,m}^{\text{TDD-1}}+R_{d,T,m}^{\text{TDD-1}}\right)\Big)
\end{align}
Furthermore, the approximated lower bounded achievable ergodic rate of TDD systems can be derived by following nearly the same steps as above for the MDD systems, except only the following two points. The first one is that all the channels within one frame are predicted relying on the first $\tau_p$ training symbols sent at the start of the frame, yielding $\pmb{\delta}(\tau_p,\alpha)$ in \eqref{eq:MDD-CEF:Rgyd1} being changed to $[\alpha^{i-\tau_p},\alpha^{i+1-\tau_p},...,\alpha^{i-1}]$. The second point is that the TDD systems do not experience SI. Therefore, we have $\pmb{\Xi}_{\text{SI}}[i] = 0, \forall i$ in \eqref{eq:MDD-CEF:XiSI}.

\subsection{Type \uppercase\expandafter{\romannumeral 2} Frame Structure}

The ergodic achievable rate of the MDD systems with the Type \uppercase\expandafter{\romannumeral 2} frame structure shown in Fig. \ref{figure-MDDCEF-FrameStructure_DUL} can be expressed as 
\begin{align}\label{eq:MDD-CEF:MDD2LB}
&R^{\text{MDD-2}}(\tau_p.\kappa_p)=\frac{1}{TM_\text{sum}}\sum\limits_{d=1}^D\Big(\sum\limits_{i=\kappa_p+1}^{\tau_p}\sum\limits_{m=1}^MR_{d,i,m}^{\text{MDD-1}} \nonumber \\
&+\sum\limits_{i=\tau_p+1}^{T}\sum\limits_{m=1}^MR_{d,i,m}^{\text{MDD-2-DL}}+\sum\limits_{i=\tau_p+1}^{T}\sum\limits_{\bar{m}=1}^{\bar{M}}R_{d,i,\bar{m}}^{\text{MDD-2-UL}}\Big)
\end{align}
where $R_{d,i,m}^{\text{MDD-1}}$ denotes the rate attained during Phase \uppercase\expandafter{\romannumeral 1}, $R_{d,i,m}^{\text{MDD-2-DL}}$ and $R_{d,i,\bar{m}}^{\text{MDD-2-UL}}$ denote respectively the DL and UL rates achievable during Phase \uppercase\expandafter{\romannumeral 2}. To compute the lower bounded rate $R^{\text{MDD-2-LB}}(\tau_p.\kappa_p)$, in \eqref{eq:MDD-CEF:MDD2LB}, the lower bound for $R_{d,i,m}^{\text{MDD-1}}$ is given by \eqref{eq:MDD-CEF:MDD1LB}, while the lower bound for $R_{d,i,m}^{\text{MDD-2-DL}}$ can be obtained from the approach for computing $R_{d,i,m}^{\text{MDD-1-LB}}$ by replacing $\check{h}_{d,i,m}^{\text{WP}}$ using $\check{h}_{d,i,m}^{\text{DD}}$. Therefore, we only need to derive the lower bound for the achievable UL ergodic rate, i.e., $R_{d,i,\bar{m}}^{\text{MDD-2-UL}} $. Denote the effective channel gain as $\omega_{d,d}^{\text{MRC}}[i,\bar{m}]=\check{\pmb{w}}_d^H[i,\bar{m}]\pmb{h}_d[i,\bar{m}]$. Then, we have \eqref{eq:MDD-CEF:MDD2ULLB} at the top of next page, 
\begin{figure*}[!t]
\hrulefill
\begin{align} \label{eq:MDD-CEF:MDD2ULLB}
R_{d,i,\bar{m}}^{\text{MDD-2-UL}} \geq R_{d,i,\bar{m}}^{\text{MDD-2-UL-LB}}&=\text{log}_2\Big(1+\frac{p_{\text{UL}}\left|\mathbb{E}\left[\omega_{d,d}^{\text{MRC}}[i,\bar{m}]\right]\right|^2}{p_{\text{UL}}\sum\limits_{k=1,k\neq d}^{D}\mathbb{E}\left[\left|\omega_{d,k}^{\text{MRC}}\left[i,\bar{m}\right]\right|^2\right]+\left(\text{var}\left\{\pmb{z}^{\text{SI}}[i]\right\}+\sigma^2\right)\mathbb{E}\left[\left\|\check{\pmb{w}}_d[i,\bar{m}]\right\|_2^2\right]}\Big)  \nonumber\\
&\overset{(a)}{=}\text{log}_2\Big(1+\frac{p_{\text{UL}}\left|\mathbb{E}\left[\check{\pmb{h}}_d^{\text{DD},H}[i,\bar{m}]\pmb{h}_d[i,\bar{m}]\right]\right|^2}{p_{\text{UL}}\sum\limits_{k=1,k\neq d}^{D}\mathbb{E}\left[\left|\check{\pmb{h}}_d^{\text{DD},H}[i,\bar{m}]\pmb{h}_k[i,\bar{m}]\right|^2\right]+\left(\xi_{\text{BS}}p_{\text{DL}}M+\sigma^2\right)\mathbb{E}\left[\left\|\check{\pmb{h}}_d^{\text{DD}}[i,\bar{m}]\right\|_2^2\right]}\Big)
\end{align}
\vspace*{-6pt}
\end{figure*}
where (a) is obtained by using $\check{\pmb{w}}_d[i,\bar{m}]=\check{\pmb{h}}_d^{\text{DD}}[i,\bar{m}]$ to implement the MRC receiver. In particular, the expectation term in the numerator of \eqref{eq:MDD-CEF:MDD2ULLB} can be derived as
\begin{align}\label{eq:MDD-CEF:MDD2ULLBsig}
&\left|\mathbb{E}\left[\check{\pmb{h}}_d^{\text{DD},H}[i,\bar{m}]\pmb{h}_d[i,\bar{m}]\right]\right|^2 \nonumber \\
&=\left|\mathbb{E}\left[\check{\pmb{h}}_d^{\text{DD},H}[i,\bar{m}]\left(\check{\pmb{h}}_d^{\text{DD}}[i,\bar{m}]+\check{\pmb{v}}_{d}^{\text{DD}}[i,\bar{m}]\right)\right]\right|^2 \nonumber \\
&\overset{(b)}{=}\left|\mathbb{E}\left[\check{\pmb{h}}_d^{\text{DD},H}[i,\bar{m}]\check{\pmb{h}}_d^{\text{DD}}[i,\bar{m}]\right]\right|^2 \nonumber \\
&=\mathbb{E}\left[\left\|\check{\pmb{h}}_d^{\text{DD}}[i,\bar{m}]\right\|_2^4\right]
\end{align}
where (b) is due to the fact that $\check{\pmb{v}}_{d}^{\text{DD}}[i,m]$ has zero mean and is independent of $\check{\pmb{h}}_d[i,\bar{m}]$. The first expectation in the denominator of \eqref{eq:MDD-CEF:MDD2ULLB} can be derived as
\begin{align}\label{eq:MDD-CEF:MDD2ULLBinter}
&\mathbb{E}\left[\left|\check{\pmb{h}}_d^{\text{DD},H}[i,\bar{m}]\pmb{h}_k[i,\bar{m}]\right|^2\right] \nonumber \\
&=\mathbb{E}\left[\left|\check{\pmb{h}}_d^{\text{DD},H}[i,\bar{m}]\check{\pmb{h}}_k^{\text{DD}}[i,\bar{m}]+\check{\pmb{h}}_d^{\text{DD},H}[i,\bar{m}]\check{\pmb{v}}_{k}^{\text{DD}}[i,\bar{m}]\right|^2\right] \nonumber \\
&\overset{(c)}{\leq}\mathbb{E}\left[\left|\check{\pmb{h}}_d^{\text{DD},H}[i,\bar{m}]\check{\pmb{h}}_k^{\text{DD}}[i,\bar{m}]\right|^2\right]+\mathbb{E}\left[\left|\check{\pmb{h}}_d^{\text{DD},H}[i,\bar{m}]\check{\pmb{v}}_{k}^{\text{DD}}[i,\bar{m}]\right|^2\right] \nonumber \\
&\overset{(d)}{=}R_h^k\mathbb{E}\left[\left\|\check{\pmb{h}}_d^{\text{DD}}[i,\bar{m}]\right\|_2^2\right]
\end{align}
where (c) and (d) follow from the fact that $\check{\pmb{h}}_d^{\text{DD}}[i,\bar{m}]$, $\check{\pmb{h}}_k^{\text{DD}}[i,\bar{m}]$ and $\check{\pmb{v}}_k^{\text{DD}}[i,\bar{m}]$ are independent vectors and $\check{\pmb{v}}_k^{\text{DD}}[i,\bar{m}]$ has zero mean. Consequently, upon substituting \eqref{eq:MDD:CEF:hnDDOr}, \eqref{eq:MDD-CEF:MDD2ULLBsig} and \eqref{eq:MDD-CEF:MDD2ULLBinter} into \eqref{eq:MDD-CEF:MDD2ULLB}, we can obtain the approximated lower-bounded UL rate, which is given by
\begin{equation}\label{eq:MDD-CEF:MDD2ULLB_2}
R_{d,i,\bar{m}}^{\text{MDD-2-UL-LB}}\approx\text{log}_2\Big(1+\frac{p_{\text{UL}}N\left(\pmb{\Theta}_i[\bar{m}]\right)_{d,d}}{p_{\text{UL}}\sum\limits_{k=1,k\neq d}^{D}\frac{\beta_k}{M_{\text{sum}}}+\xi_{\text{BS}}p_{\text{DL}}M+\sigma^2}\Big)
\end{equation}

In the context of TDD, the achievable ergodic rate in the case of the Type \uppercase\expandafter{\romannumeral 2} frame structure shown in Fig. \ref{figure-MDDCEF-FrameStructure_DUL} can be expressed as
\begin{align}\label{eq:MDD-CEF:TDD2LB}
R^{\text{TDD-2}}(\tau_p,\tau_u)&=\frac{1}{TM_\text{sum}}\sum\limits_{d=1}^D\bigg(\sum\limits_{i=\tau_p+1}^{\tau_p+\tau_u}\sum\limits_{m=1}^{M_{\text{sum}}}R_{d,i,m}^{\text{TDD-2-UL}} \nonumber \\
&+\sum\limits_{i=\tau_p+\tau_u+2}^{T}\sum\limits_{m=1}^{M_{\text{sum}}}R_{d,i,m}^{\text{TDD-2-DL}} \nonumber \\
&+ \frac{1}{2}\sum\limits_{m=1}^{M_{\text{sum}}}\left(R_{d,\tau_p+\tau_u+1,m}^{\text{TDD-2-UL}}+R_{d,T,m}^{\text{TDD-2-DL}}\right)\bigg)
\end{align}
Based on \eqref{eq:MDD-CEF:TDD2LB}, the approximated lower bound $R^{\text{TDD-2-LB}}(\tau_p,\tau_u)$ can be obtained in the way similar to that for MDD systems. Therefore, the derivation is omitted to avoid redundancy. 

\begin{table}
\caption{Simulation parameters}
\centering
\begin{tabular}{|l|l|}
\hline
Default parameters & Value  \\ \hline
Distance between BS and MTs ($D_d$) &  $[50,100]\ m$  \\ \hline
Number of BS antennas ($N$)  & 32  \\ \hline
Number of MTs ($D$) & 8 \\ \hline
Number of DL/UL subcarriers for TDD ($M_{\text{sum}}$) & 96 \\ \hline
Number of DL/UL subcarriers for MDD ($M/\bar{M}$) & 64 / 32 \\ \hline
BS transmit power ($p_{\text{DL}}M$) & $30 \ \text{dBm}$ \\ \hline
MT transmit power ($p_{\text{UL}}\bar{M}$) & $20 \ \text{dBm}$  \\ \hline
Noise power  & -94 dBm   \\ \hline
Large-scale fading ($\beta_d$) & $(D_d)^{3.8}$ \\ \hline 
Delay taps ($L$) & 4 \\ \hline 
Carrier center frequency & 5 GHz \\ \hline
Subcarrier bandwidth & 15 kHz  \\ \hline
OFDM symbol duration ($T_s$) & 66.67 $\mu s$ \\ \hline
Total symbols in one frame ($T$) & $28$  \\ \hline
Switching interval in TDD & 0.5 \\ \hline
Digital modulation & 16 QAM  \\ \hline 
SIC capability at BS ($\xi_{\text{BS}}$) & 130 dB \\ \hline 
SIC capability at MT ($\xi_{\text{MT}}$) & 120 dB \\ \hline 
\end{tabular}
\label{Table:MDD-CEF:para}
\end{table}

\section{Simulation Results and Discussion}\label{sec:MDD-CEF:Sim}

In this section, we present the numerical results for the performance comparison of the MDD and TDD systems with two types of frame structures, as shown in Fig. \ref{figure-MDDCEF-FrameStructure_DL} and Fig. \ref{figure-MDDCEF-FrameStructure_DUL}, respectively. 
We also compare the performance of IBFD and MDD systems with the proposed frame structures, when SIC capability is limited. Furthermore, we validate the analytical results derived in Section \ref{sec:MDD-CEF:AER} by Monte Carlo simulations over 1000 channel realizations. Throughout this section, we assume a single-cell network with one BS and multiple single-antenna MTs. Some key parameters are listed in Table \ref{Table:MDD-CEF:para}. It is noteworthy that to demonstrate the tendency of the rate change within one frame, the switching intervals in the TDD mode are not considered, while the X-axis represents the index of OFDM symbol. By contrast, the switching intervals are taken into account in evaluating the average sum rate over one frame, as shown in Eqs. \eqref{eq:MDD-CEF:TDD1LB} and \eqref{eq:MDD-CEF:TDD2LB}.

\subsection{Type \uppercase\expandafter{\romannumeral 1} Frame Structure: TDD Vs MDD} \label{sec:MDD-CEF:Sim:sub1}

In this subsection, TDD and MDD systems are compared under the Type \uppercase\expandafter{\romannumeral 1} frame structure. For TDD systems, a total of 7 pilots are transmitted within one frame. By contrast, for MDD systems, UL pilots can always be transmitted or be transmitted in part. Note that, in the following figures, `TDD-1' denotes the scheme that all pilots are consecutively transmitted at the beginning of the frame and a $7$-th order WP is applied. The notations of `TDD-1, Ideal' and `TDD-1, NoP' indicate respectively that the channel is ideal without the channel aging problem and that the CSI acquisition does not rely on channel prediction. `MDD-1(z)' denotes the scheme that UL pilots are continuously sent by MTs and a $z$-th order WP is applied. Furthermore, the results of `TDD-1' and `MDD-1(z)' are plotted by using the lower-bounded rates given in \eqref{eq:MDD-CEF:MDD1kp}, \eqref{eq:MDD-CEF:TDD1LB}, \eqref{eq:MDD-CEF:MDD2LB} and \eqref{eq:MDD-CEF:TDD2LB}, while the results of `TDD-1, Apx' and `MDD-1(z), Apx' represent the correspondingly approximated rates derived in Section \ref{sec:MDD-CEF:AER}.

\begin{figure}
\centering
\includegraphics[width=0.9\linewidth]{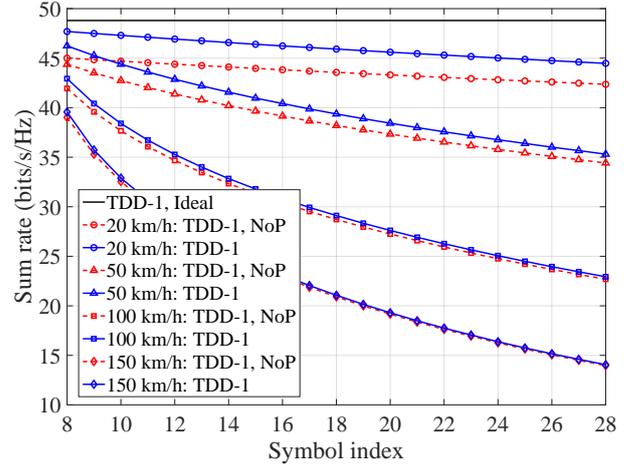} 
\caption{Sum rate versus OFDM symbol index of the TDD systems with the different assumptions about channel acquisition and relative velocity.}
\label{figure-MDDCEF-fig0}
\end{figure}

In Fig. \ref{figure-MDDCEF-fig0}, we evaluate the performance of the TDD systems operated in different mobility scenarios with or without channel prediction. We observe that when the relative velocity is 20 km/h, the TDD system with WP outperforms the TDD system without channel prediction. However, as the symbol index increases, meaning that CSI becomes more outdated, the performance gap between the above two strategies becomes narrower. Therefore, in the TDD mode, the latter symbols benefit less from channel prediction, as only 7 symbols are used for prediction and the prediction error increases over time. Additionally, as seen in Fig. \ref{figure-MDDCEF-fig0}, as the relative speed increases, the advantage of employing channel prediction diminishes, and the performance drops faster and becomes significantly worse than the system with precise CSI. The results in Fig. \ref{figure-MDDCEF-fig0} explicitly indicate that the traditional TDD mode is not feasible for communication in high-mobility scenarios, even when channel prediction is employed.

\begin{figure}[]
\centering
\subfigure[Normalized MSE]{
\includegraphics[width=0.9\linewidth]{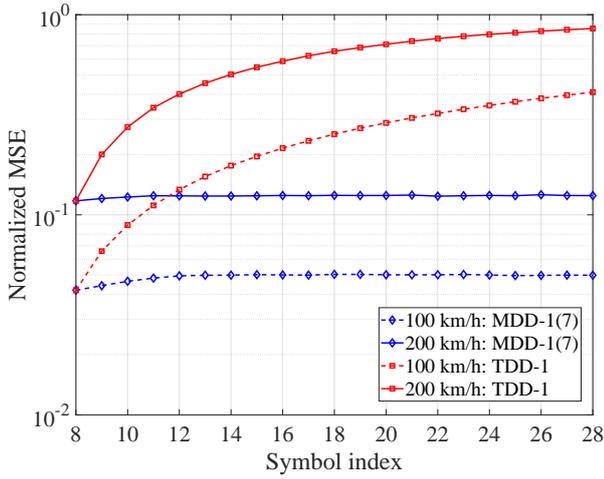}
}
\subfigure[Sum rate]{
\includegraphics[width=0.9\linewidth]{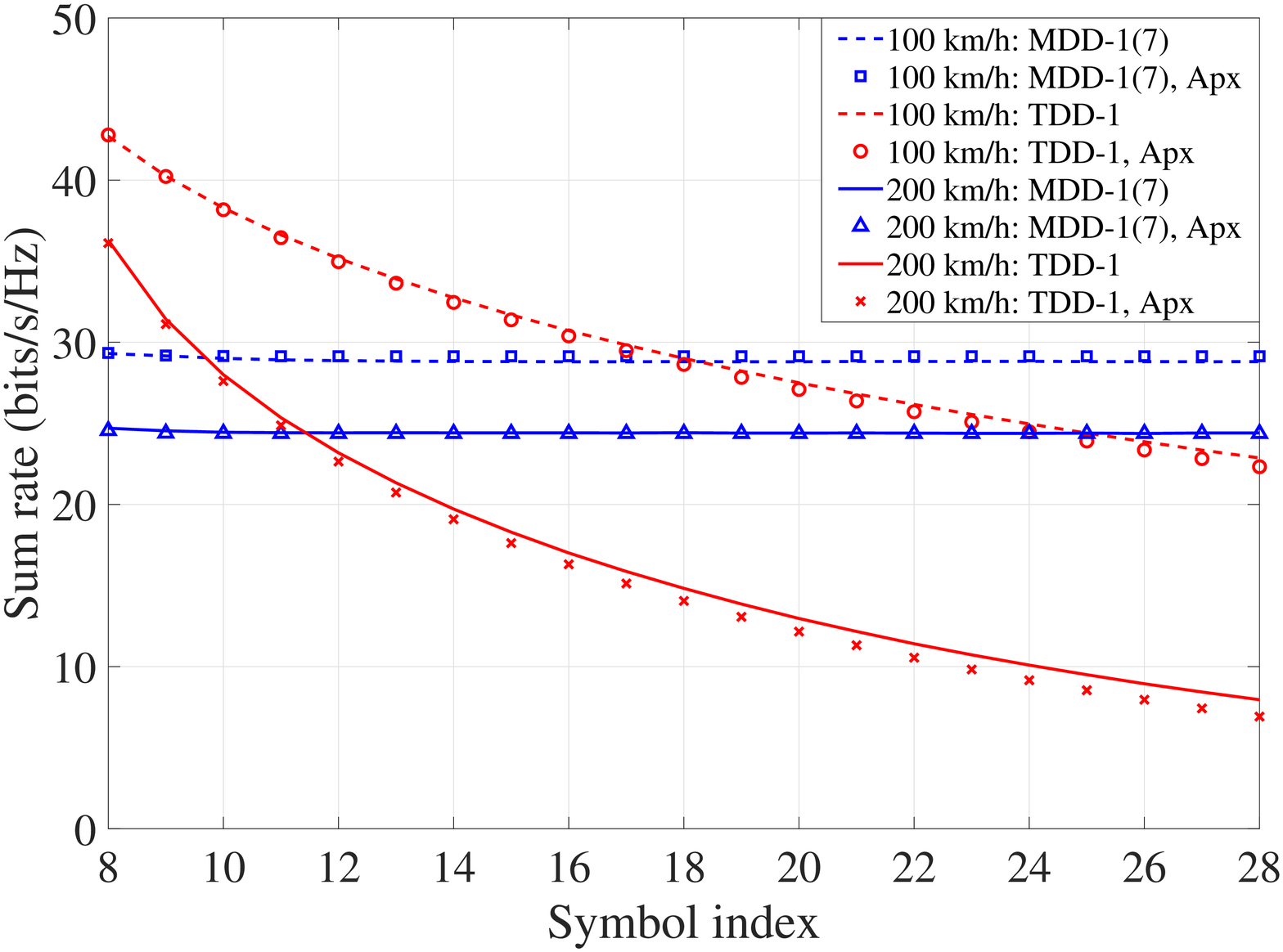}
}
\caption{Performance comparison of the MDD and TDD systems, when Type \uppercase\expandafter{\romannumeral 1} frame structure and 7-th order WP are employed.}
\label{figure-MDDCEF-fig1}
\end{figure}

Based on the Type \uppercase\expandafter{\romannumeral 1} frame structure presented in Fig. \ref{figure-MDDCEF-FrameStructure_DL}, we demonstrate the effect of the time-varying channels on the performance of TDD and MDD systems in Fig. \ref{figure-MDDCEF-fig1}. It can be observed from Fig. \ref{figure-MDDCEF-fig1}(a) that the accuracy of channel prediction in MDD systems is nearly constant after 7 pilot symbols, 
  regardless of the relative traveling velocity of 100 km/h or 200 km/h. This is because in MDD systems, the UL pilots can continuously provide BS with the most updated CSI for DL transmission, which hence guarantees a stable system sum rate, as shown in Fig. \ref{figure-MDDCEF-fig1}(b). Moreover, the difference between the sum rates achieved by the MDD systems at 100 km/h and 200 km/h is relatively small, meaning that MDD systems are robust to time-varying fading channels. By contrast, although the sum rate of TDD systems is higher than that of MDD systems within the first several symbols, for both the velocities considered, it reduces quickly with time in terms of the OFDM symbol indices, due to the reduced accuracy of channel prediction with time. Furthermore, when the velocity is changed from 100 km/h to 200 km/h, the sum rate deteriorates significantly. Therefore, the performance of TDD systems is sensitive to the mobility of wireless channels. The rationale is that in TDD systems, UL training and DL have to be performed in an alternative way. Once it is switched to DL transmission, the CSI has to be predicted based on the old data, which may be outdated, if mobility is high. Additionally, in Fig. \ref{figure-MDDCEF-fig1}, it is demonstrated that the approximations of the lower bounded ergodic rates derived in Section \ref{sec:MDD-CEF:AER} match closely to the lower bounded sum rates attained via Monte Carlo simulations. 

\begin{figure}[]
\centering
\subfigure[Sum rate versus OFDM symbol index.]{
\includegraphics[width=0.9\linewidth]{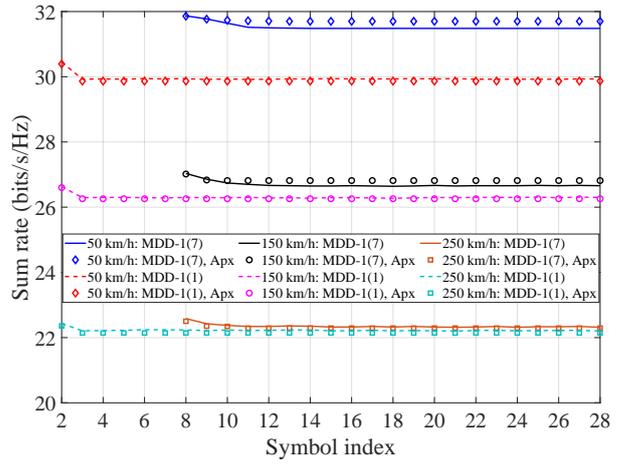}
}
\subfigure[Average sum rate within one frame versus relative velocity.]{
\includegraphics[width=0.9\linewidth]{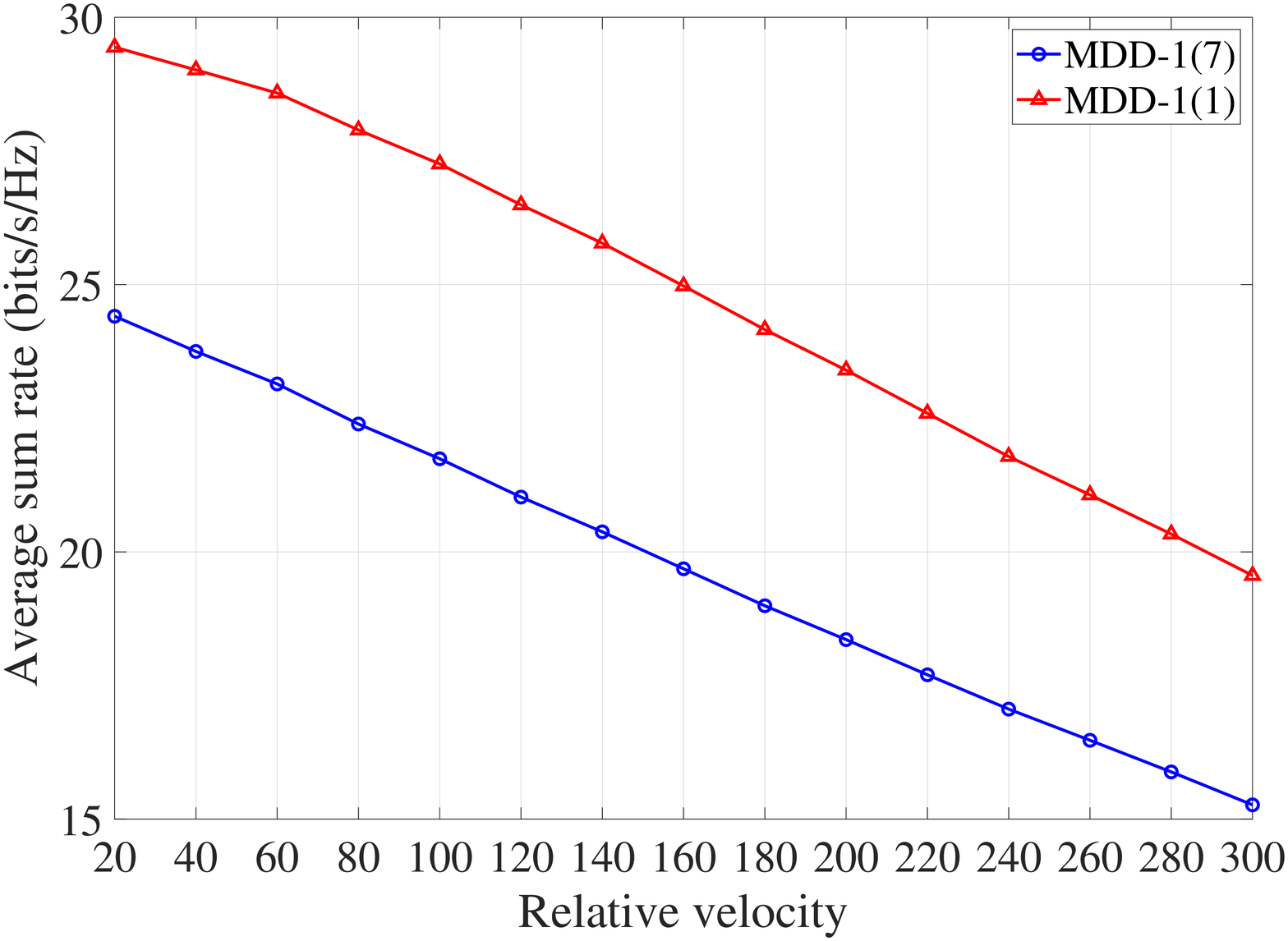}
}
\caption{Performance comparison of the MDD systems with Type \uppercase\expandafter{\romannumeral 1} frame structure and different orders of WPs.}
\label{figure-MDDCEF-fig2}
\end{figure}

\begin{figure}[]
\centering
\subfigure[Evenly spaced pilots in TDD Type \uppercase\expandafter{\romannumeral 1} frame structure.]{
\includegraphics[width=0.9\linewidth]{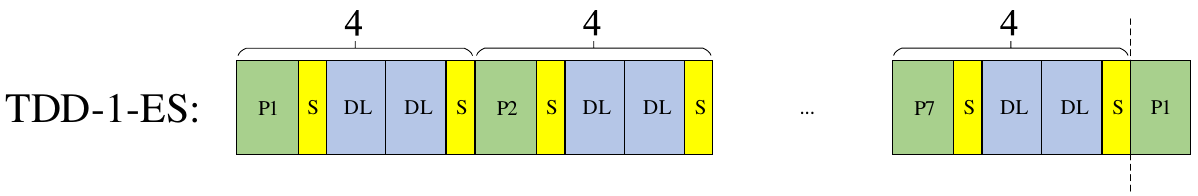}
}
\subfigure[Two groups of pilots in TDD Type \uppercase\expandafter{\romannumeral 1} frame structure.]{
\includegraphics[width=0.9\linewidth]{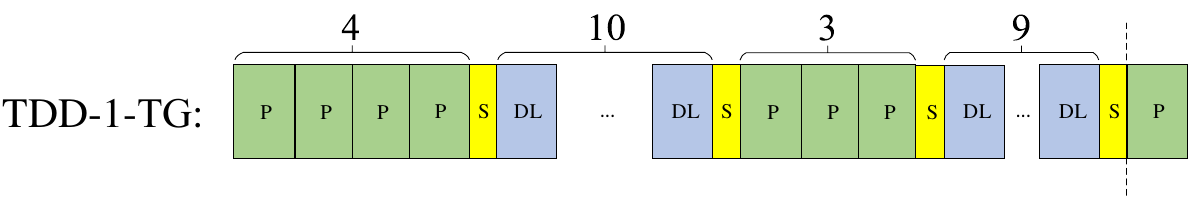}
}
\caption{Alternative pilot distribution methods for the Type \uppercase\expandafter{\romannumeral 1} frame structure in TDD mode, when the total number of pilots is 7.}
\label{figure-MDDCEF-figES}
\end{figure}

\begin{figure}[t]
\centering
\subfigure[$T=28$]{
\includegraphics[width=0.9\linewidth]{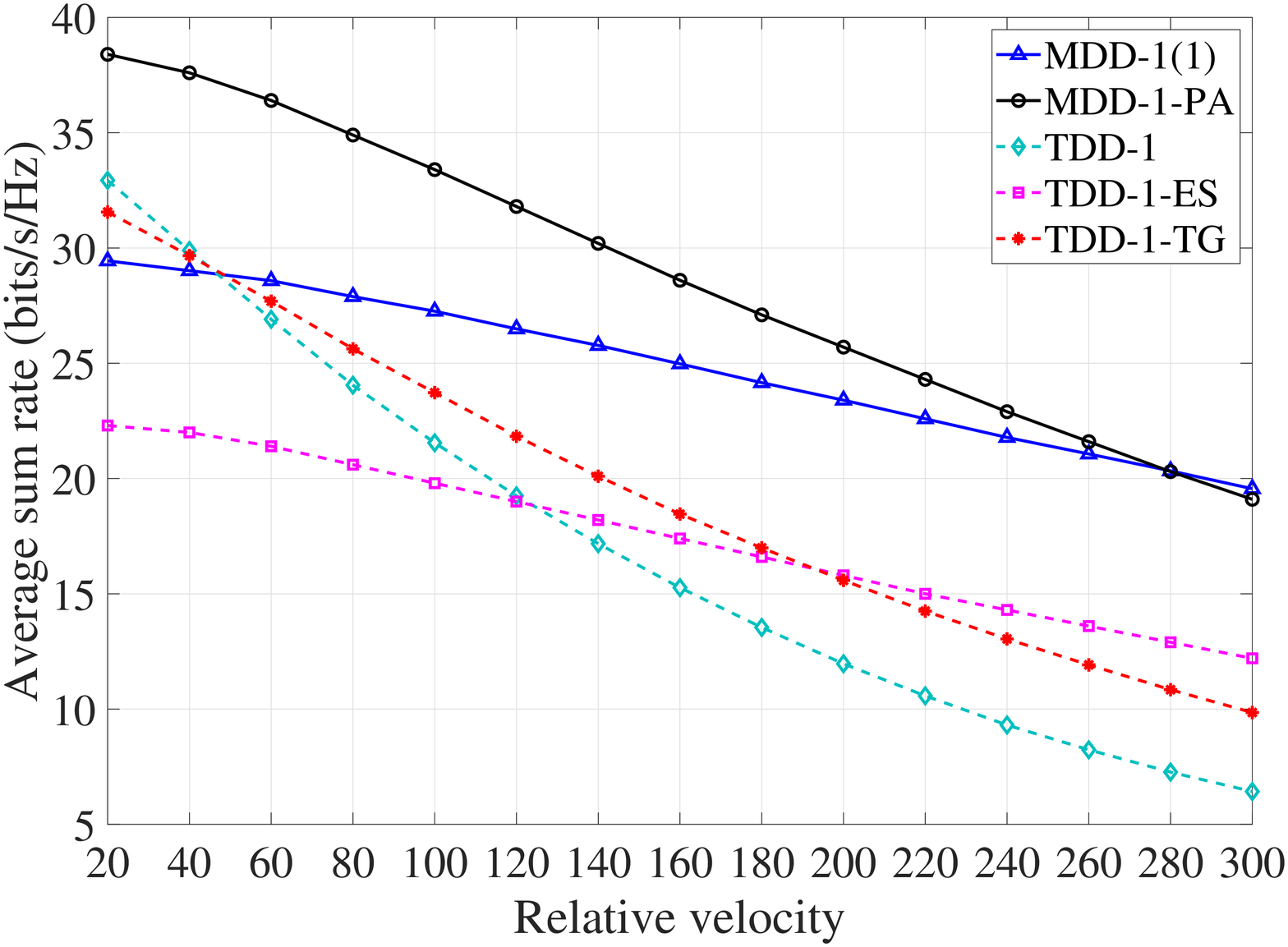}
}
\subfigure[$T=56$]{
\includegraphics[width=0.9\linewidth]{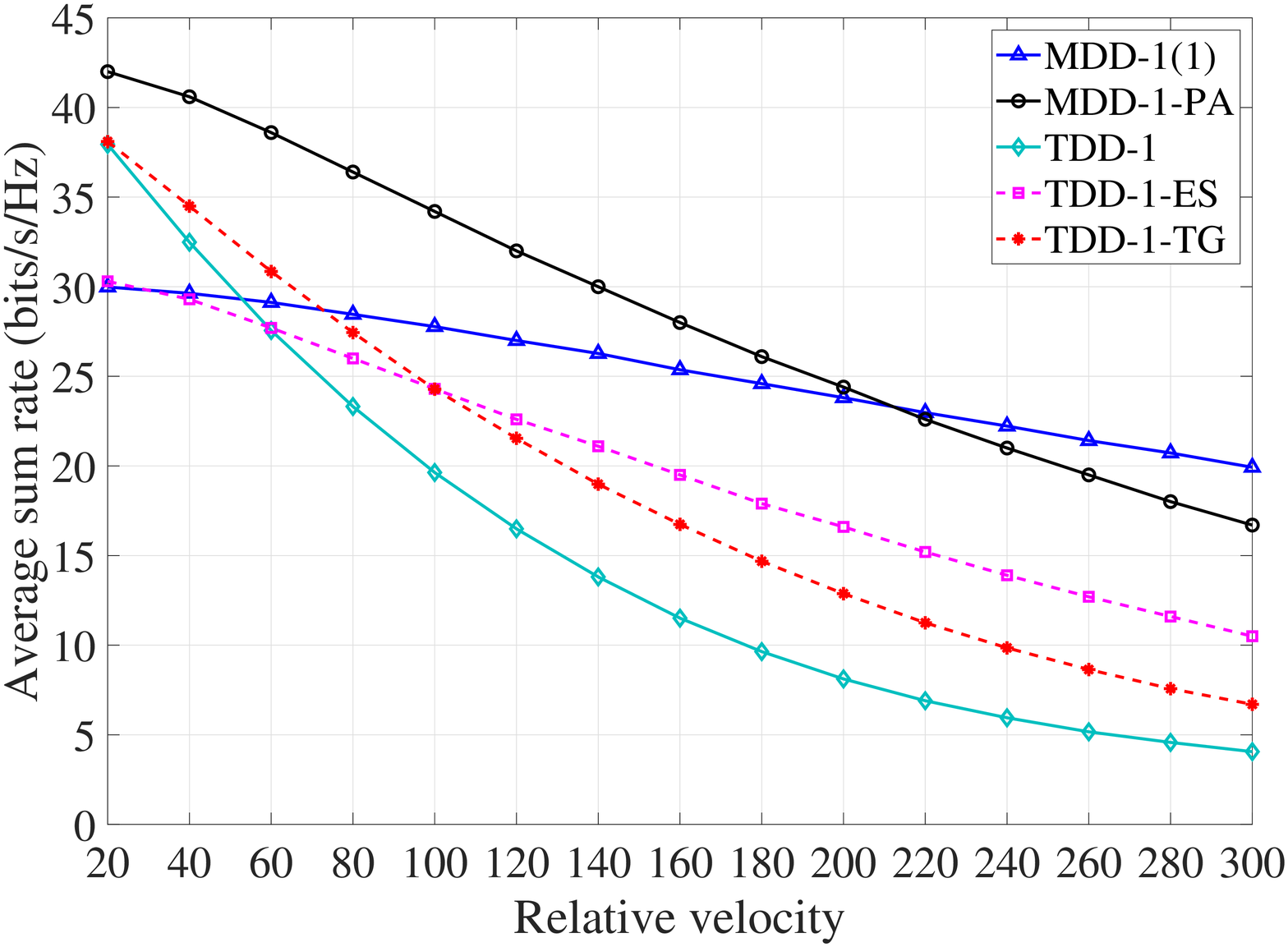}
}
\caption{Average sum rate versus relative velocity, when the Type \uppercase\expandafter{\romannumeral 1} frame structure is used.}
\label{figure-MDDCEF-fig4}
\end{figure}

In Fig. \ref{figure-MDDCEF-fig2}, we investigate the performance of MDD systems, when Type \uppercase\expandafter{\romannumeral 1} frame structure and different orders of WPs are employed. According to Fig. \ref{figure-MDDCEF-FrameStructure_DL}, a higher order WP uses more channel observations received in the past to predict channels. It is observed from Fig. \ref{figure-MDDCEF-fig2}(a) that at the relative speed of 50 km/h, the $7$-th order WP in general significantly outperforms the 1-st order WP. However, when the speed is increased to 250 km/h, the 1-st order WP achieves nearly the same performance as the $7$-th order WP. The reason behind is that when channel varies fast, the prediction of the current channel is more dependent on the nearest pilot symbols and the observations in the past become less correlated to the prediction. Fig. \ref{figure-MDDCEF-fig2}(b) further shows that when pilots are continuously transmitted under the Type \uppercase\expandafter{\romannumeral 1} frame structure, the $1$-st order WP attains a better performance averaged over one frame, owing to the fact that in this case, 6 more DL symbols can be transmitted, even though the slightly reduced prediction accuracy results in some performance loss.

Next, to comprehensively compare the performance of the TDD and MDD systems employing the Type \uppercase\expandafter{\romannumeral 1} frame, we consider two more pilot distribution methods for TDD, namely the TDD-1-ES and TDD-1-TG, as shown in Fig. \ref{figure-MDDCEF-figES}. Specifically, with the TDD-1-ES, 7 training symbols are evenly distributed within one frame. By contrast, with the TDD-1-TG, 7 training symbols are first divided into two groups, which are then inserted into the frame. Note that the TDD-1-TG is similar to the frame structure applied in 3GPP 36.211, where two pilot subframes are evenly distributed within one frame \cite{access2009physical}. For MDD systems, we also introduce an alternative Type \uppercase\expandafter{\romannumeral 1} frame structure denoted as MDD-1-PA, as depicted in Fig. \ref{figure-MDDCEF-FrameStructure_DL}. With the MDD-1-PA, $\tau_p=1$, and a total of 7 pilots are activated, which are evenly distributed within one frame. 

In Fig. \ref{figure-MDDCEF-fig4}, we compare the average sum rate of the MDD and TDD systems with the various Type \uppercase\expandafter{\romannumeral 1} frame structures, and when two frame lengths are considered. Note that, as the frame length increases, the number of pilots is set to remain the same in MDD-1-PA and three TDD structures, while the number of DL symbols increases. It can be observed from Fig. \ref{figure-MDDCEF-fig4} that MDD-1-PA outperforms all the other arrangements, when the relative speed is lower than 280 km/h in Fig.~\ref{figure-MDDCEF-fig4}(a) and 210 km/h in Fig.~\ref{figure-MDDCEF-fig4}(b). This is because when only a fraction of pilots are transmitted, the saved UL resources (subcarriers) can be exploited to transmit DL data,  provided that the CSI updating is enough for protecting the DL transmission. By contrast, MDD-1(1) is superior to the MDD-1-PA in the high-speed case. In this case, although there are only $M$ subcarriers used for DL transmission, the continuously transmitted pilots on the $\bar{M}$ UL subcarriers guarantees the reliable channel prediction, which the MDD-1-PA structure is  however unable to provide. With respect to TDD, the advantages of the three frame structures are dependent on the relative velocity and the length of frames. As shown in Fig. \ref{figure-MDDCEF-fig4}(a) and \ref{figure-MDDCEF-fig4}(b), in general, TDD-1 and TDD-TG have relatively better performance, when the relative velocity is low, owing to the use of less switching intervals. By contrast, TDD-1-ES benefits from the frequent transmission of training symbols within one frame and hence performs better in the high-speed long-frame case. 


\begin{figure}[]
\centering
\subfigure[Normalized MSE]{
\includegraphics[width=0.9\linewidth]{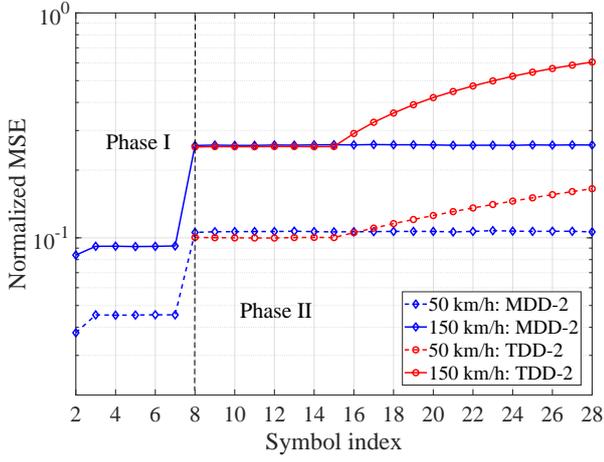}
}
\subfigure[Sum rate]{
\includegraphics[width=0.9\linewidth]{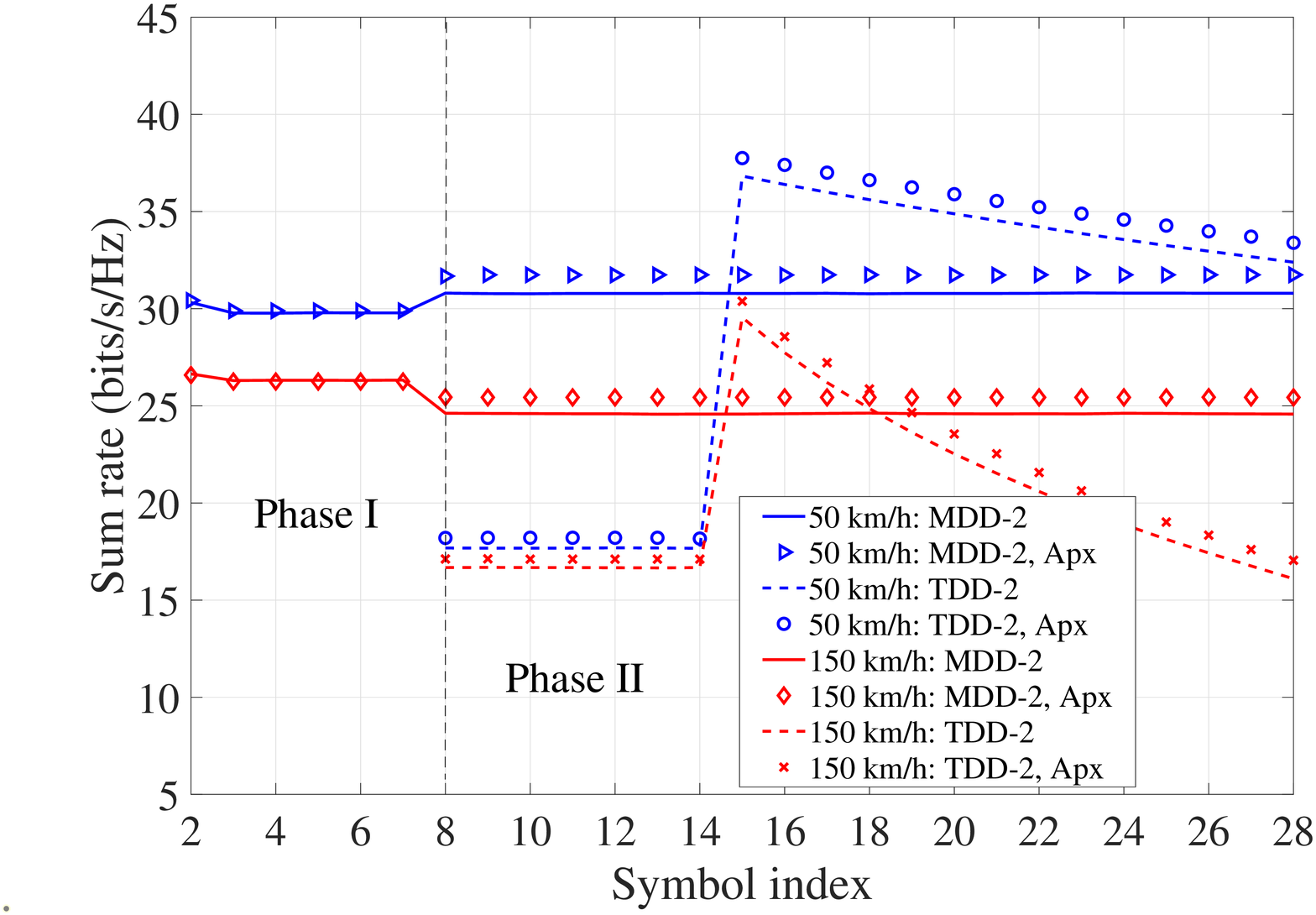}
}
\caption{Performance comparison of the MDD and TDD systems, when Type \uppercase\expandafter{\romannumeral 2} frame structure is applied.}
\label{figure-MDDCEF-fig3}
\end{figure}

\begin{figure}
\centering
\includegraphics[width=0.9\linewidth]{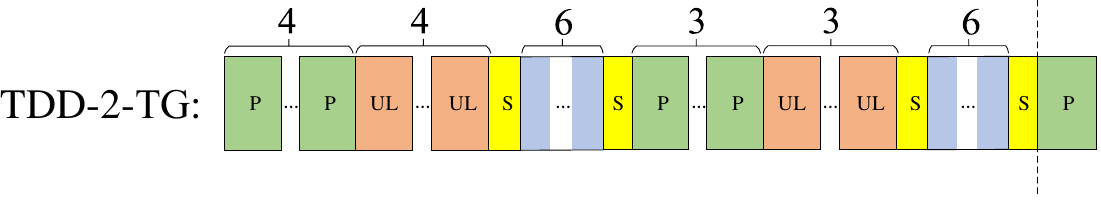}
\caption{Distribution of two groups of pilots in TDD Type \uppercase\expandafter{\romannumeral 2} frame structure, when the total number of pilots is 7.}
\label{figure-MDDCEF-figES2}
\end{figure}

\subsection{Type \uppercase\expandafter{\romannumeral 2} Frame Structure: TDD Vs MDD}

In this subsection, we compare the performance of MDD and TDD systems operated with the Type \uppercase\expandafter{\romannumeral 2} frame structure. In the following figures, `TDD-2' and `MDD-2' denote respectively the TDD and MDD systems employing the proposed frame structure as shown in Fig. \ref{figure-MDDCEF-FrameStructure_DUL} with $\tau_p=7$, $\tau_u=7$ and $\kappa_p=1$.

The performance comparison of MDD and TDD with respect to symbol index is plotted in Fig. \ref{figure-MDDCEF-fig3}, when Type \uppercase\expandafter{\romannumeral 2} frame structure is considered. In conjunction with Fig. \ref{figure-MDDCEF-FrameStructure_DUL}, we can see that, as the MDD system is able to operate DL and UL simultaneously during Phase \uppercase\expandafter{\romannumeral 1}, DL data symbols can be transmitted while collecting the channel observations for prediction in Phase \uppercase\expandafter{\romannumeral 2}.From Fig. \ref{figure-MDDCEF-fig3}(a) we observe that the WP in Phase I yields less prediction errors when compared with the DD-WP in Phase II. This is as expected, since the pilots used for WP in Phase I are known to BS. In Phase \uppercase\expandafter{\romannumeral 2}, although the performance of channel prediction is better without the effect of SI, TDD system only activates the UL transmission from the 8-th to the 14-th symbol, leading to lower sum rate when compared with the MDD system, which is capable of transmitting data over both DL and UL. At the 15-th symbol, both MDD and TDD systems apply the $7$-th DD-WP and the TDD system attains a higher sum rate, as it uses all the subcarriers for DL transmission. However, similar to the Type \uppercase\expandafter{\romannumeral 1} frame structure, new UL data symbols are not available for the TDD system to update channel causing the degraded accuracy of channel prediction. Consequently, its sum rate decreases, as symbol index increases.

\begin{figure}[]
\centering
\subfigure[$T=28$]{
\includegraphics[width=0.9\linewidth]{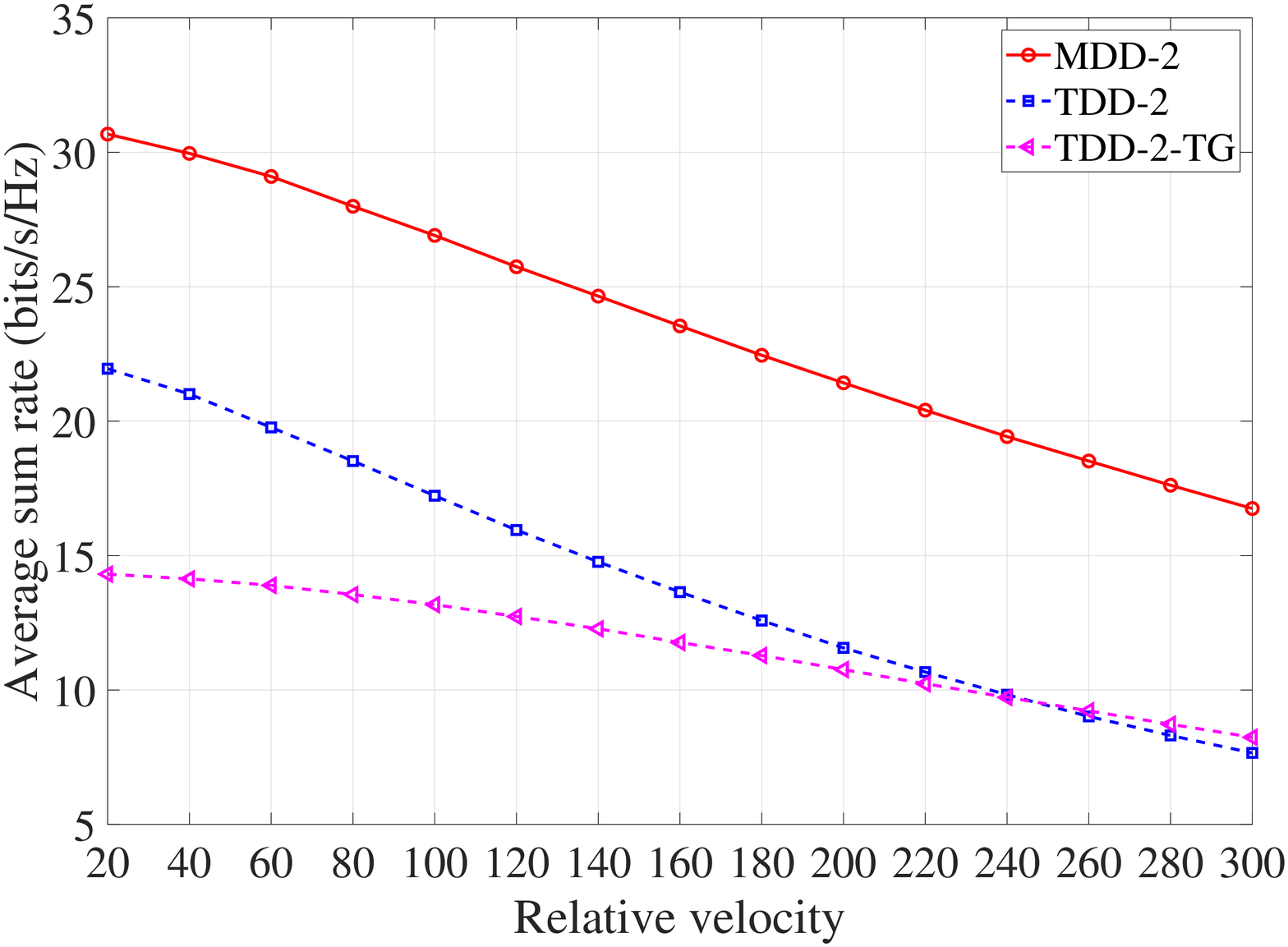}
}
\subfigure[$T=56$]{
\includegraphics[width=0.9\linewidth]{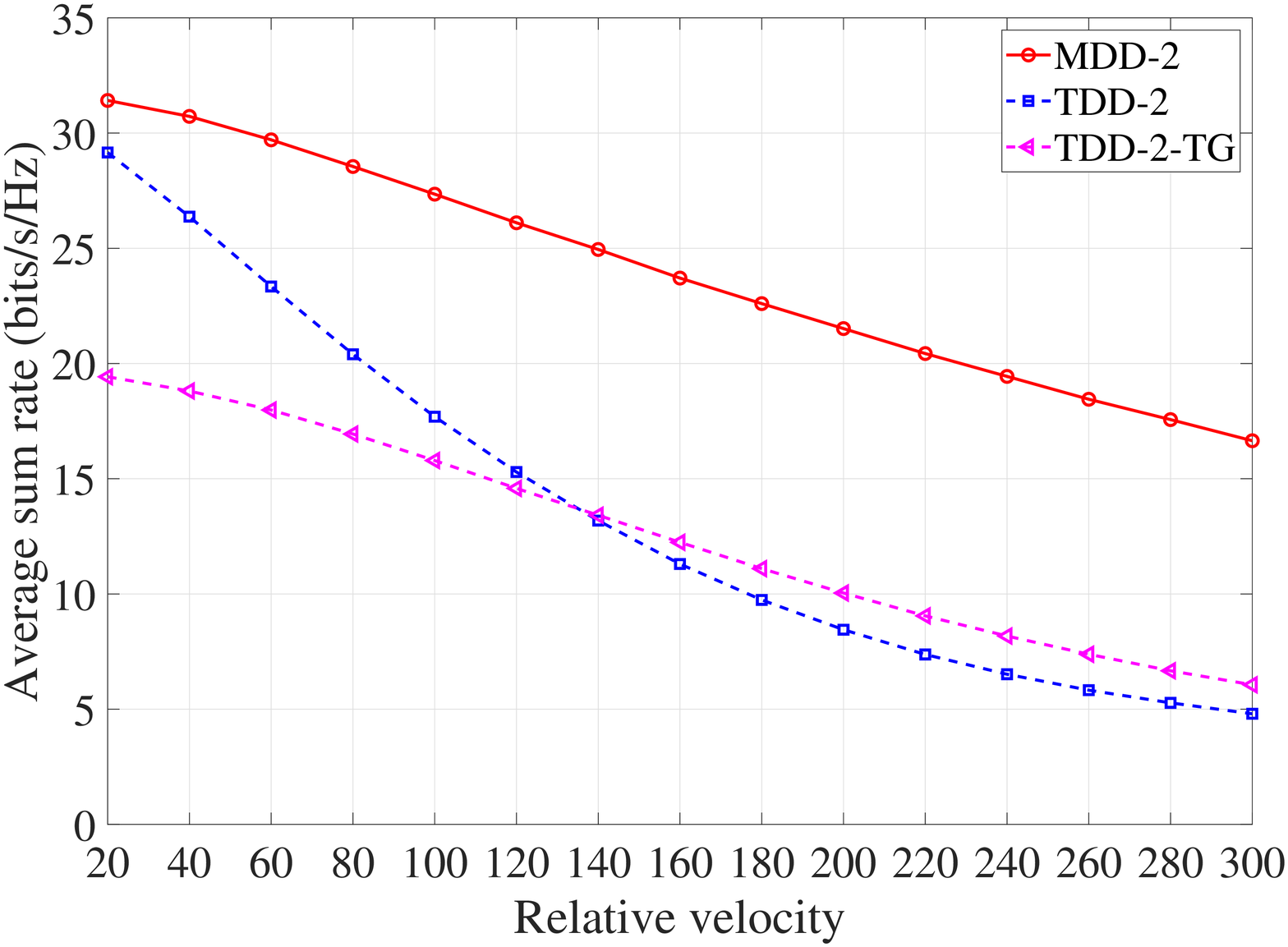}
}
\caption{Average sum rate versus the relative velocity, when the Type \uppercase\expandafter{\romannumeral 2} frame structure is used.}
\label{figure-MDDCEF-fig5}
\end{figure}

\begin{figure}
\centering
\includegraphics[width=0.9\linewidth]{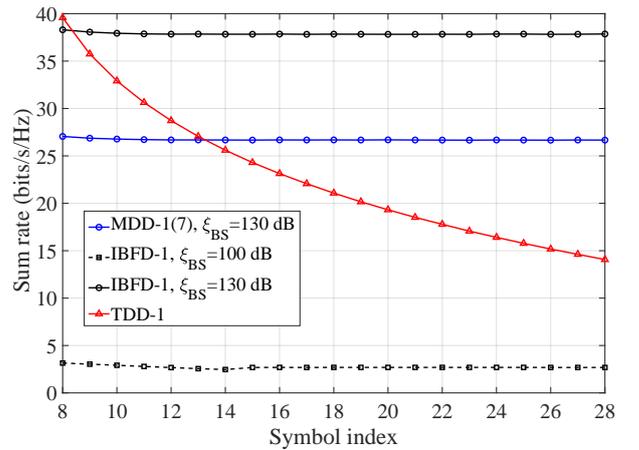} 
\caption{Sum rate comparison of MDD, IBFD and TDD with respect to OFDM symbol index, when Type \uppercase\expandafter{\romannumeral 1} frame structure and the relative velocity of 150 km/h are assumed.}
\label{figure-MDDCEF-fig6}
\end{figure}

In Fig. \ref{figure-MDDCEF-fig5}, we compare the average sum rate versus the relative speed of the MDD and TDD systems, when the Type \uppercase\expandafter{\romannumeral 2} frame structure is employed. Here, we also consider an optional frame structure for TDD systems, which is shown in Fig. \ref{figure-MDDCEF-figES2}, where two groups of pilots are separately inserted into one frame. We can observe from Fig. \ref{figure-MDDCEF-fig5} that owing to the DL transmission occurred in both Phase \uppercase\expandafter{\romannumeral 1} and the start of Phase \uppercase\expandafter{\romannumeral 2}, MDD significantly outperforms TDD over the whole velocity, ranging from 20 km/h to 300 km/h, for both the frame-length considered. Specifically, for the TDD in low-mobility case, the rate achieved by the TDD-2-TG is considerably lower than that achieved by the TDD-2. However, benefiting from the extra training in the middle of the frame, the TDD-2-TG finally surpasses the TDD-2 at 240 km/h when $T=28$, and at 140 km/h when $T=56$. The results in Fig. \ref{figure-MDDCEF-fig5} imply that in the general communication scenarios where UL training, UL transmission and DL transmissions are all required at the same time, MDD shows significant advantages over TDD, when the channel aging problem is encountered.

\subsection{Influence of SIC on IBFD and MDD}

In this part, we demonstrate the impact of SIC on the performance of MDD and IBFD systems. To this end, we consider the conventional IBFD with the Type \uppercase\expandafter{\romannumeral 1} frame structure denoted as `IBFD-1', where UL pilots are unremittingly transmitted over $\bar{M}$ subcarriers and a $7$-th order WP is applied for channel prediction. Note that, unlike the MDD-1 that only uses $M$ subcarriers for DL transmission, as stated in Section \ref{sec:MDD-CEF:Sim:sub1}, IBFD-1 uses all the $M_\text{sum}$ subcarriers for the DL transmission, which imposes an extra SIC burden in digital-domain. According to \cite{bharadia2013full}, the digital-domain SIC in IBFD systems is expected to cancel the main linear signal component by more than 30 dB. By contrast, MDD systems are almost free from linear SI in the digital-domain owing to the FFT operation, which separates the UL subcarriers from the DL subcarriers. Based on this observation, it is sensible to assume that with the same amount of system resources used for SI suppression, MDD can always achieve 30 dB more SIC than IBFD. From Fig. \ref{figure-MDDCEF-fig6}, it is shown that when both IBFD and MDD systems have the same SI mitigation capability\footnote{This means IBFD costs more computation effort and system complexity than MDD for SIC, and hence achieves lower energy efficient \cite{li2021multicarrier}.}, IBFD apparently outperforms both TDD and MDD. However, if IBFD can only mitigate the SI by 100 dB\footnote{Note that 100~dB of SI is often assumed in IBFD systems, which is regarded as the bottleneck value based on the existing SIC approaches~\cite{di2016joint,ng2016power}.}, the achievable sum rate of IBFD systems drops dramatically, and is much lower than that of MDD systems.

\section{Conclusions}\label{sec:MDD-CEF:con}

A MDD-assisted mMIMO has been studied and comprehensively compared with the TDD-relied mMIMO over fast time-varying channels. Two types of MDD frame structures incorporated respectively with WP and DD-WP have been proposed to combat channel aging. For comparison, the frames in TDD mMIMO have been structured according to the 3GPP standards or improved for handling the channel aging problem. MDD has also been compared with the IBFD by considering their SIC capabilities. Moreover, the approximated lower bounds for the achievable rates of the MDD and TDD mMIMO systems communicating based on the proposed frame structures have been derived. According to our studies and performance comparison, we have the following main observations. First, when the Type \uppercase\expandafter{\romannumeral 1} frame structure is employed, benefiting from the instantly available CSI updating while without involving UL/DL switching, MDD outperforms TDD over time-varying channels, which is more significant when mobility becomes higher. Second, the above performance advantage of MDD over TDD retains under the Type \uppercase\expandafter{\romannumeral 2} frame structure, owing to the MDD's merits that it can proactively start DL transmission in Phase \uppercase\expandafter{\romannumeral 1}, while simultaneously updating the CSI with the aid of the UL transmission in Phase \uppercase\expandafter{\romannumeral 2}. Third, it is shown that SIC is critical to IBFD systems. By taking the advantage of near-free SI in the digital-domain, MDD is capable of achieving better performance than IBFD in the presence of channel aging, if SIC in IBFD systems is imperfect. Additionally, the approximated lower bounded rates derived in this paper have been validated by the Monte-Carlo simulations, showing that they agree well with each other.

\appendices

\section{Calculation of SI Power}\label{Appen:MDD:CEF-1}
\subsubsection{SI in Equation \eqref{eq:MDD-CEF:ydDetect}}
According to the model for residual SI \cite{day2012full2}, the covariance of $z_d^{\text{SI}}[i]$ is 
\begin{align}\label{eq:MDD-CEF:Varzd}
\text{cov}\left\{z_d^{\text{SI}}[i]\right\}&=\mathbb{E}\left[h_{\text{SI}}x_d[i]x_d^H[i]h_{\text{SI}}^H\right] \nonumber \\
 &=\xi_{\text{MT}}p_{\text{UL}}\bar{M}
\end{align}
when using $\mathbb{E}\left[h_{\text{SI}}h_{\text{SI}}^H\right]=1$. \eqref{eq:MDD-CEF:Varzd} is obtained by substituting $x_d[i]=\sqrt{p_{\text{UL}}}\sum\limits_{\bar{m}=1}^{\bar{M}}x_d[i,\bar{m}]$.

\subsubsection{SI in Equation \eqref{eq:MDD-CEF:sUL}}
To compute the covariance of $z^{\text{SI}}[i]$ in \eqref{eq:MDD-CEF:sUL},  we again refer to the residual model in \cite{day2012full2} and have 
\begin{align}\label{eq:MDD-CEF:covzSIi}
\text{cov}\left\{\pmb{z}^{\text{SI}}[i]\right\}=\xi_{\text{BS}}\text{diag}\left(\mathbb{E}\left[\pmb{H}_{\text{SI}}\pmb{s}_{\text{DL}}[i]\pmb{s}_{\text{DL}}^H[i]\pmb{H}_{\text{SI}}^H\right]\right)
\end{align}
Upon substituting $\pmb{s}_{\text{DL}}[i]=\sum\limits_{m=1}^M\sqrt{p_{\text{DL}}}\pmb{F}^{\text{ZF}}[i,m]\pmb{x}[i,m]$ into \eqref{eq:MDD-CEF:covzSIi}, we can obtain the SI power at the $n$-th receive element as
\begin{align}
&\text{cov}\left\{\pmb{z}^{\text{SI}}[i]\right\}_{n,n} =\xi_{\text{BS}} \mathbb{E}\left[\pmb{H}_{\text{SI}}^{(n,:)}\pmb{s}_{\text{DL}}[i]\pmb{s}_{\text{DL}}^H[i]\pmb{H}_{\text{SI}}^{(n,:),H}\right] \nonumber \\
&=\xi_{\text{BS}}\mathbb{E}\left[\text{Tr}\left(p_{\text{DL}}\sum\limits_{m=1}^M\pmb{H}_{\text{SI}}^{(n,:)}\pmb{F}^{\text{ZF}}[i,m]\pmb{F}^{\text{ZF},H}[i,m]\pmb{H}_{\text{SI}}^{(n,:),H}\right)\right] \nonumber \\
&\overset{(a)}{=}\xi_{\text{BS}}p_{\text{DL}}\sum\limits_{m=1}^M\left\|\pmb{F}^{\text{ZF}}[i,m]\right\|_F^2\nonumber \\
&\overset{(b)}{=}\xi_{\text{BS}}p_{\text{DL}}M
\end{align}
where $(a)$ is obtained using $\mathbb{E}[\pmb{H}_{\text{SI}}^{(n,:),H}\pmb{H}_{\text{SI}}^{(n,:)}]=\pmb{I}_N$ and $(b)$ is due to the power normalization $\left\|\pmb{F}^{\text{ZF}}[i,m]\right\|_F^2=1$. Consequently, $\text{cov}\left\{\pmb{z}^{\text{SI}}[i]\right\}=\xi_{\text{BS}}p_{\text{DL}}M\pmb{I}_N$. 

\section{Covariance matrix of predicted UL subcarrier channels in \eqref{eq:MDD:CEF:gndDDPre}}\label{Appen:MDD:CEF-2}
The covariance matrix $\pmb{\Gamma}_{d,i}$ in \eqref{eq:MDD:CEF:gndDDPre} can be written as \eqref{eq:MDD-CEF:AppenGamma}.
\begin{figure*}[!t]
\hrulefill
\begin{align}\label{eq:MDD-CEF:AppenGamma}
\pmb{\Gamma}_{d,i}&=\mathbb{E}\left[\check{\pmb{h}}_{n,d}^{\text{UL}}[i+1]\check{\pmb{h}}_{n,d}^{\text{UL},H}[i+1]\right] =\left[
\begin{array}{ccc}
\mathbb{E}\left[\check{h}_{n,d}^{\text{DD}}[i+1,1]\check{h}_{n,d}^{\text{DD},H}[i+1,1]\right] & \cdots & \mathbb{E}\left[\check{h}_{n,d}^{\text{DD}}[i+1,1]\check{h}_{n,d}^{\text{DD},H}[i+1,\bar{M}]\right] \\
\vdots  & \ddots & \vdots \\
\mathbb{E}\left[\check{h}_{n,d}^{\text{DD}}[i+1,\bar{M}]\check{h}_{n,d}^{\text{DD},H}[i+1,1]\right]& \cdots &\mathbb{E}\left[\check{h}_{n,d}^{\text{DD}}[i+1,\bar{M}]\check{h}_{n,d}^{\text{DD},H}[i+1,\bar{M}]\right] 
\end{array}\right] 
\end{align}
\vspace*{-6pt}
\end{figure*}
As for  the diagonal elements in $\pmb{\Gamma}_{d,i}$, according to \eqref{eq:MDD:CEF:hnDDOr}, we have
\begin{equation}
h_{n,d}[i+1,\bar{m}]=\check{h}_{n,d}^{\text{DD}}[i+1,\bar{m}]+\check{e}^{\text{DD}}_{n,d}[i+1,\bar{m}]
\end{equation}
Hence, the variance of $\check{h}_{n,d}^{\text{DD}}[i+1,\bar{m}]$ is $\left(\pmb{\Theta}_i[\bar{m}]\right)_{d,d}$, given in \eqref{eq:MDD:CEF:hnDDOr}. For the off-diagonal elements of $\pmb{\Gamma}_{d,i}$, we have
\begin{align}
&\mathbb{E}\left[\check{h}_{n,d}^{\text{DD}}[i+1,\bar{m}_1]\check{h}_{n,d}^{\text{DD},H}[i+1,\bar{m}_2]\right] \nonumber \\
&= \mathbb{E}\Big[(h_{n,d}[i+1,\bar{m}_1]-\check{e}_{n,d}^{\text{DD}}[i+1,\bar{m}_1])(h_{n,d}^{H}[i+1,\bar{m}_2] \nonumber \\
&-\check{e}_{n,d}^{\text{DD},H}[i+1,\bar{m}_2])\Big] \nonumber \\
&\overset{(a)}{=}\mathbb{E}\left[h_{n,d}[i+1,\bar{m}_1]h_{n,d}^{H}[i+1,\bar{m}_2]\right] \nonumber \\
&\overset{(b)}{=}\mathbb{E}\left[\pmb{\psi}_{\bar{m}_1}\pmb{g}_{n,d}[i+1]\pmb{g}^H_{n,d}[i+1]\pmb{\psi}^H_{\bar{m}_2}\right] \nonumber \\
&=\frac{\beta_d}{L}\pmb{\psi}_{\bar{m}_1}\pmb{\psi}^H_{\bar{m}_2}, \ \bar{m}_1\neq\bar{m}_2
\end{align}
where (a) is obtained based on the fact that the prediction error of the $\bar{m}_1$-th subcarrier channel is independent of the $\bar{m}_2$-th subcarrier channel and vice versa, while (b) can be derived from the FFT operation. In summary, the covariance matrix $\pmb{\Gamma}_{d,i}$ can be expressed as
\begin{align}
\pmb{\Gamma}_{d,i}=\left[
\begin{array}{ccc}
\left(\pmb{\Theta}_i[1]\right)_{d,d} & \cdots & \frac{\beta_d}{L}\pmb{\psi}_{1}\pmb{\psi}^H_{\bar{M}} \\
\vdots  & \ddots & \vdots \\
\frac{\beta_d}{L}\pmb{\psi}_{\bar{M}}\pmb{\psi}^H_{1}& \cdots &\left(\pmb{\Theta}_i[\bar{M}]\right)_{d,d} 
\end{array}\right] 
\end{align}




\ifCLASSOPTIONcaptionsoff
  \newpage
\fi



%

\bibliographystyle{IEEEtran}
\bibliography{MDD-CEF}

\begin{thebibliography}{10}
\providecommand{\url}[1]{#1}
\csname url@samestyle\endcsname
\providecommand{\newblock}{\relax}
\providecommand{\bibinfo}[2]{#2}
\providecommand{\BIBentrySTDinterwordspacing}{\spaceskip=0pt\relax}
\providecommand{\BIBentryALTinterwordstretchfactor}{4}
\providecommand{\BIBentryALTinterwordspacing}{\spaceskip=\fontdimen2\font plus
\BIBentryALTinterwordstretchfactor\fontdimen3\font minus
  \fontdimen4\font\relax}
\providecommand{\BIBforeignlanguage}[2]{{%
\expandafter\ifx\csname l@#1\endcsname\relax
\typeout{** WARNING: IEEEtran.bst: No hyphenation pattern has been}%
\typeout{** loaded for the language `#1'. Using the pattern for}%
\typeout{** the default language instead.}%
\else
\language=\csname l@#1\endcsname
\fi
#2}}
\providecommand{\BIBdecl}{\relax}
\BIBdecl

\bibitem{larsson2013massive}
E.~G. Larsson, O.~Edfors, F.~Tufvesson, and T.~L. Marzetta, ``Massive {MIMO}
  for next generation wireless systems,'' \emph{arXiv preprint
  arXiv:1304.6690}, 2013.

\bibitem{bjornson2016massive}
E.~Bj{\"o}rnson, E.~G. Larsson, and T.~L. Marzetta, ``Massive {{MIMO}}: {{Ten}}
  myths and one critical question,'' \emph{IEEE Communications Magazine},
  vol.~54, no.~2, pp. 114--123, 2016.

\bibitem{sanguinetti2019toward}
L.~Sanguinetti, E.~Bj{\"o}rnson, and J.~Hoydis, ``Toward massive {{MIMO}} 2.0:
  Understanding spatial correlation, interference suppression, and pilot
  contamination,'' \emph{IEEE Transactions on Communications}, vol.~68, no.~1,
  pp. 232--257, 2019.

\bibitem{truong2013effects}
K.~T. Truong and R.~W. Heath, ``Effects of channel aging in massive {{MIMO}}
  systems,'' \emph{Journal of Communications and Networks}, vol.~15, no.~4, pp.
  338--351, 2013.

\bibitem{papazafeiropoulos2015deterministic}
A.~K. Papazafeiropoulos and T.~Ratnarajah, ``Deterministic equivalent
  performance analysis of time-varying massive {MIMO} systems,'' \emph{IEEE
  Transactions on Wireless Communications}, vol.~14, no.~10, pp. 5795--5809,
  2015.

\bibitem{kashyap2017performance}
S.~Kashyap, C.~Moll{\'e}n, E.~Bj{\"o}rnson, and E.~G. Larsson, ``Performance
  analysis of ({{TDD}}) massive {MIMO} with kalman channel prediction,'' in
  \emph{2017 IEEE International Conference on Acoustics, Speech and Signal
  Processing (ICASSP)}.\hskip 1em plus 0.5em minus 0.4em\relax IEEE, 2017, pp.
  3554--3558.

\bibitem{yuan2020machine}
J.~Yuan, H.~Q. Ngo, and M.~Matthaiou, ``Machine learning-based channel
  prediction in massive {{MIMO}} with channel aging,'' \emph{IEEE Transactions
  on Wireless Communications}, vol.~19, no.~5, pp. 2960--2973, 2020.

\bibitem{mirza2018performance}
J.~Mirza, G.~Zheng, K.-K. Wong, S.~Lambotharan, and L.~Hanzo, ``On the
  performance of multiuser {MIMO} systems relying on full-duplex {CSI}
  acquisition,'' \emph{IEEE Transactions on Communications}, vol.~66, no.~10,
  pp. 4563--4577, 2018.

\bibitem{rajashekar2019multicarrier}
R.~Rajashekar, C.~Xu, N.~Ishikawa, L.-L. Yang, and L.~Hanzo, ``Multicarrier
  division duplex aided millimeter wave communications,'' \emph{IEEE Access},
  vol.~7, pp. 100\,719--100\,732, 2019.

\bibitem{yang2009multicarrier}
L.-L. Yang, \emph{Multicarrier Communications}.\hskip 1em plus 0.5em minus
  0.4em\relax John Wiley \& Sons, 2009.

\bibitem{bharadia2013full}
D.~Bharadia, E.~McMilin, and S.~Katti, ``Full duplex radios,'' in \emph{ACM
  SIGCOMM Computer Communication Review}, vol.~43, no.~4.\hskip 1em plus 0.5em
  minus 0.4em\relax ACM, 2013, pp. 375--386.

\bibitem{li2020self}
B.~Li, L.-L. Yang, R.~G. Maunder, and S.~Sun, ``Self-interference cancellation
  and channel estimation in multicarrier-division duplex systems with hybrid
  beamforming,'' \emph{IEEE Access}, vol.~8, pp. 160\,653--160\,669, 2020.

\bibitem{li2021resource}
B.~Li, L.-L. Yang, R.~G. Maunder, and S.~Sun, ``Resource allocation in
  millimeter-wave multicarrier-division duplex systems with hybrid
  beamforming,'' \emph{IEEE Transactions on Vehicular Technology}, vol.~70,
  no.~8, pp. 7921--7935, 2021.

\bibitem{li2021multicarrier}
B.~Li, L.-L. Yang, R.~G. Maunder, P.~Xiao, and S.~Sun, ``Multicarrier-division
  duplex: A duplexing technique for the shift to {6G} wireless
  communications,'' \emph{IEEE Vehicular Technology Magazine}, vol.~16, no.~4,
  pp. 57--67, 2021.

\bibitem{etsi2013136}
3GPP, ``Physical layer procedures ({Release} 15), {TS} 36.213.''

\bibitem{3gpp2017nr}
3GPP, ``{NR}; physical channels and modulation ({Release} 15), {TS} 38.211.''

\bibitem{nam2015joint}
C.~Nam, C.~Joo, and S.~Bahk, ``Joint subcarrier assignment and power allocation
  in full-duplex {{OFDMA}} networks,'' \emph{IEEE Transactions on Wireless
  Communications}, vol.~14, no.~6, pp. 3108--3119, 2015.

\bibitem{li2016energy}
X.~Li, X.~Ge, X.~Wang, J.~Cheng, and V.~C. Leung, ``Energy efficiency
  optimization: Joint antenna-subcarrier-power allocation in {OFDM-DASs},''
  \emph{IEEE Transactions on Wireless Communications}, vol.~15, no.~11, pp.
  7470--7483, 2016.

\bibitem{zhang2004multiuser}
Y.~J. Zhang and K.~B. Letaief, ``Multiuser adaptive subcarrier-and-bit
  allocation with adaptive cell selection for {OFDM} systems,'' \emph{IEEE
  Transactions on Wireless Communications}, vol.~3, no.~5, pp. 1566--1575,
  2004.

\bibitem{sabharwal2014band}
A.~Sabharwal, P.~Schniter, D.~Guo, D.~W. Bliss, S.~Rangarajan, and R.~Wichman,
  ``In-band full-duplex wireless: Challenges and opportunities.'' \emph{IEEE
  Journal on Selected Areas in Communications}, vol.~32, no.~9, pp. 1637--1652,
  2014.

\bibitem{kolodziej2019band}
K.~E. Kolodziej, B.~T. Perry, and J.~S. Herd, ``In-band full-duplex technology:
  Techniques and systems survey,'' \emph{IEEE Transactions on Microwave Theory
  and Techniques}, 2019.

\bibitem{baddour2005autoregressive}
K.~E. Baddour and N.~C. Beaulieu, ``Autoregressive modeling for fading channel
  simulation,'' \emph{IEEE Transactions on Wireless Communications}, vol.~4,
  no.~4, pp. 1650--1662, 2005.

\bibitem{day2012full2}
B.~P. Day, A.~R. Margetts, D.~W. Bliss, and P.~Schniter, ``Full-duplex {{MIMO}}
  relaying: Achievable rates under limited dynamic range,'' \emph{IEEE Journal
  on Selected Areas in Communications}, vol.~30, no.~8, pp. 1541--1553, 2012.

\bibitem{debaillie2014analog}
B.~Debaillie, D.-J. van~den Broek, C.~Lavin, B.~van Liempd, E.~A. Klumperink,
  C.~Palacios, J.~Craninckx, B.~Nauta, and A.~P{\"a}rssinen, ``Analog/{RF}
  solutions enabling compact full-duplex radios,'' \emph{IEEE Journal on
  Selected Areas in Communications}, vol.~32, no.~9, pp. 1662--1673, 2014.

\bibitem{kolodziej2016multitap}
K.~E. Kolodziej, J.~G. McMichael, and B.~T. Perry, ``Multitap {RF} canceller
  for in-band full-duplex wireless communications,'' \emph{IEEE Transactions on
  Wireless Communications}, vol.~15, no.~6, pp. 4321--4334, 2016.

\bibitem{haykin2005adaptive}
S.~S. Haykin, \emph{Adaptive Filter Theory}.\hskip 1em plus 0.5em minus
  0.4em\relax Pearson Education India, 2005.

\bibitem{schafhuber2005mmse}
D.~Schafhuber and G.~Matz, ``{MMSE} and adaptive prediction of time-varying
  channels for {OFDM} systems,'' \emph{IEEE Transactions on Wireless
  Communications}, vol.~4, no.~2, pp. 593--602, 2005.

\bibitem{khansefid2015achievable}
A.~Khansefid and H.~Minn, ``Achievable downlink rates of {{MRC}} and {{ZF}}
  precoders in massive {{MIMO}} with uplink and downlink pilot contamination,''
  \emph{IEEE Transactions on Communications}, vol.~63, no.~12, pp. 4849--4864,
  2015.

\bibitem{access2009physical}
3GPP, ``{Evolved Universal Terrestrial Radio Access (E-UTRA); Physical channels
  and modulation (Release 8)}, {TS} 36.211.''

\bibitem{di2016joint}
B.~Di, S.~Bayat, L.~Song, Y.~Li, and Z.~Han, ``Joint user pairing, subchannel,
  and power allocation in full-duplex multi-user {OFDMA} networks,'' \emph{IEEE
  Transactions on Wireless Communications}, vol.~15, no.~12, pp. 8260--8272,
  2016.

\bibitem{ng2016power}
D.~W.~K. Ng, Y.~Wu, and R.~Schober, ``Power efficient resource allocation for
  full-duplex radio distributed antenna networks,'' \emph{IEEE Transactions on
  Wireless Communications}, vol.~15, no.~4, pp. 2896--2911, 2016.

\end{thebibliography}

%








\end{document}